\documentclass[journal]{IEEEtran} 
\usepackage{amsfonts}  
\usepackage{amssymb}  
\usepackage{makecell} 
\usepackage{booktabs} 
\usepackage{threeparttable}

\usepackage{algorithm}
\usepackage{algpseudocode}
\usepackage{graphicx}
\usepackage{subfigure} 

\usepackage{amsthm}

\usepackage{xcolor}

\usepackage{cite}

\ifCLASSINFOpdf
   \graphicspath{{../pdf/}{../jpeg/}}
   \DeclareGraphicsExtensions{.pdf,.jpeg,.png}
\else
   \usepackage[dvips]{graphicx}
   \graphicspath{{../eps/}}
   \DeclareGraphicsExtensions{.eps}
\fi
\usepackage{amsmath}
\usepackage{array}

\ifCLASSOPTIONcompsoc
  \usepackage[caption=false,font=normalsize,labelfont=sf,textfont=sf]{subfig}
\else
  \usepackage[caption=false,font=footnotesize]{subfig}
\fi

\usepackage{stfloats}
\begin{document}
%
\title{A Relaxed Energy Function Based Analog Neural Network Approach to Target Localization in Distributed MIMO Radar}
%
%
%
\author{Xiaoyu~Zhao,~\IEEEmembership{Student~Member,~IEEE,}
	Jun~Li,~\IEEEmembership{Member,~IEEE,}
	and~Qinghua~Guo,~\IEEEmembership{Senior~Member,~IEEE}
\thanks{
	Copyright (c) 2015 IEEE. Personal use of this material is permitted. However, permission to use this material for any other purposes must be obtained from the IEEE by sending a request to pubs-permissions@ieee.org.
	This work was supported by the Fundamental Research Funds for the Central Universities under Grant XJS222601 and by the National Natural Science Foundation of China under Grant 62171350. (Corresponding authors: Jun Li and Qinghua Guo)
}
\thanks{ X. Zhao, and J. Li are with the National Laboratory of Radar Signal Processing,
	Xidian University, Xi'an 710071, China. (e-mail: zhaoxiaoyu588@126.com; junli01@mail.xidian.edu.cn).}
\thanks{Q. Guo is with the School of Electrical, Computer, and Telecommunications Engineering,
	University of Wollongong, Wollongong, NSW 2522, Australia (e-mail: qguo@uow.edu.au).}
}

\maketitle

\begin{abstract}
Analog neural networks are highly effective to solve some optimization problems, and they have been used for target localization in distributed multiple-input multiple-output (MIMO) radar. In this work, we design a new relaxed energy function based neural network (RNFNN) for target localization in distributed MIMO radar. We start with the maximum likelihood (ML) target localization with a complicated objective function, which can be transformed to a more tractable one with equality constraints by introducing some auxiliary variables. Different from the existing Lagrangian programming neural network (LPNN) methods, we further relax the optimization problem formulated for target localization, so that the Lagrangian multiplier terms are no longer needed, leading to a relaxed energy function with better convexity. Based on the relaxed energy function, a RNFNN is implemented with much simpler structure and faster convergence speed. Furthermore, the RNFNN method is extended to localization in the presence of transmitter and receiver location errors. It is shown that the performance of the proposed localization approach achieves the Cramér-Rao lower bound (CRLB) within a wider range of signal-to-noise ratios (SNRs). Extensive comparisons with the state-of-the-art approaches are provided, which demonstrate the advantages of the proposed approach in terms of performance improvement and computational complexity (or convergence speed).
\end{abstract}

\begin{IEEEkeywords}
Analog neural network, energy function, multiple-input multiple-output (MIMO) radar, target localization, antenna position errors
\end{IEEEkeywords}

%

\section{Introduction} 
\IEEEPARstart{R}{esearch} on multiple-input multiple-output (MIMO)
radar systems has been drawing significant attention because of their remarkable advantages in enhancing detection performance and improving parameter estimation accuracy recently \cite{2008-Xu-Targetdetectionparameter,2016-Khomchuk-Performanceanalysistarget,2010-He-NoncoherentMIMORadar,2021-Sadeghi-TargetLocalizationGeometry}. The essence of MIMO radar is to use multiple transmitting antennas for sending multiple linearly independent probing signals and multiple antennas for receiving the reflected target echo, and it has the ability to jointly process the signals received at multiple receiving antennas.
Basically, MIMO radars can be categorized, based on the antennas geometry, into colocated and distributed configurations \cite{2007-Li-MIMORadarColocated}, \cite{2008-Haimovich-MIMORadarWidely}. In the former, all antennas are closely spaced, and the waveform diversity can be utilized to enhance parameter identification \cite{2007-Li-ParameterIdentifiabilityMIMO,2021-Shi-ParameterIdentifiabilityDiversity} and waveform design flexibility \cite{2019-Cheng-JointDesignTransmit}.
On the other hand, the latter with widely separated antennas takes advantage of the spatial diversity of the target’s radar cross section (RCS), leading to improved target detection and localization performance \cite{2006-Fishler-Spatialdiversityradars,2010-Godrich-TargetLocalizationAccuracy}. In this work, we focus on target localization with the latter configuration.

In recent years, there has been a increasing interest in target localization problem in distributed MIMO radar 
\cite{2010-Godrich-TargetLocalizationAccuracy,2012-Niu-TargetLocalizationTracking,2014-Rui-EllipticLocalizationPerformance,2009-Godrich-Targetlocalisationtechniques,2013-Dianat-TargetLocalizationusing,2015-Einemo-Weightedleastsquares,2015-Noroozi-TargetLocalizationBistatic,2017-Amiri-AsymptoticallyEfficientTarget,2020-Noroozi-ClosedFormSolution,2017-Noroozi-IterativeTargetLocalization,2016-Liang-LagrangeProgrammingNeural,2020SunMovingTargetLocalization,2020SongApproximatelyEfficientEstimator,2021SongTargetLocalizationClock}. 
Generally speaking, target localization in such a system can be achieved in a direct or indirect manner. In direct approaches, such as \cite{2010-Godrich-TargetLocalizationAccuracy,2012-Niu-TargetLocalizationTracking,2014-Rui-EllipticLocalizationPerformance}, the target location is directly determined by collecting all the received signals and searching over a grid. These direct methods often require enormous computational power and may be impractical due to their multi-dimensional grid search and large number of grid points. 
On the other hand, indirect methods first measure the time delays (TDs) to produce the corresponding bistatic ranges (BRs), which are range sum of transmitter-target and target-receiver, multiplied with the speed of light. Then a set of elliptic equations constructed from the BR measurements are solved to obtain the target location. In \cite{2009-Godrich-Targetlocalisationtechniques}, a best linear
unbiased estimator (BLUE) was presented by linearizing the elliptic equations via Taylor series expansion. The main drawback of the BLUE method is that an initial guess sufficiently close to the target location is required. In \cite{2013-Dianat-TargetLocalizationusing}, by pair-wise subtraction of the elliptic ones two set of hyperbolic equations were derived, from which the target location can be estimated. However, elliptic positioning can usually achieve better localization accuracy than hyperbolic localization \cite{2014-Rui-EllipticLocalizationPerformance}. In \cite{2015-Einemo-Weightedleastsquares}, a quadratically constrained quadratic program (QCQP) problem was formulated after linearization of the BR measurements. By solving the QCQP with the weighted least squares (WLS) technique, target position estimation is achieved. 
Based on the construction of pseudo-linear equation set by introducing the auxiliary parameters and ignoring the noise terms of second order, various WLS algorithms have been proposed in \cite{2015-Noroozi-TargetLocalizationBistatic,2017-Amiri-AsymptoticallyEfficientTarget,2020-Noroozi-ClosedFormSolution,2017-Noroozi-IterativeTargetLocalization}. Although these WLS estimators can provide optimum performance in low noise levels, such a linearization processing, however, makes these estimators sensitive to the noise level so that in some cases the deviation from the Cramér-Rao lower bound (CRLB) occurs dramatically. In \cite{2020SunMovingTargetLocalization}, the target position estimation is achieved without knowing the noise variances, in which, however, the authors exploited some prior knowledge about the structure of the noise covariance matrix. In addition, some efficient solutions for target positioning in MIMO radars by considering the sensor location errors and time synchronization errors are respectively developed in \cite{2020SongApproximatelyEfficientEstimator} and \cite{2021SongTargetLocalizationClock}.

Analog neural networks have been shown to be powerful optimization techniques with applications in a variety of areas such as kinematic control of redundant manipulators \cite{2018-Zhang-AdaptiveProjectionNeural,2020-Li-CooperativeKinematicControl}, model predictive control \cite{2014-Yan-RobustModelPredictive,2016-Yan-TubeBasedRobust}, sparse signal reconstruction \cite{2016-Liu-$L_1$MinimizationAlgorithms,2019-Xu-DiscreteTimeProjection}, associative memory \cite{2017-Kobayashi-SymmetricComplexValued,2016-Minemoto-RetrievalperformanceHopfield}, wireless sensor network positioning \cite{2013-Li-dynamicneuralnetwork,2018-Han-AugmentedLagrangeProgramminga}, etc. In particularly, they have been shown to be highly effective in solving the problem of target localization in distributed MIMO radar \cite{2016-Liang-LagrangeProgrammingNeural,2020-Shi-RobustMIMORadar}. Based on the BR measurements, maximum likelihood (ML) target localization can be formulated, and to solve the problem, some investigations based on the Lagrangian programming neural network (LPNN) were carried out \cite{2016-Liang-LagrangeProgrammingNeural}. Although the existing LPNN method is shown good performance under the case of equal BR measurement noise variances, it exhibits considerably degraded performance in the more general non-identical noise variance cases. We note that, in practice the noise level of measured BRs depends on the signal-to-noise ratio (SNR) at each transmitter-receiver pair \cite{1984-Barton-HandbookRadarMeasurement,2005-Willis-BistaticRadar}, and they typically have different variances.
Moreover, the convergence of the Lagrangian function is relatively slow, as the update of the Lagrangian multiplier is an ascent iteration and it can only converge moderately quickly \cite{2013-Zhang-Matrixanalysisapplications}. 

In order to overcome disadvantages of the existing LPNN approach and other localization methods, we design a novel relaxed energy function based neural network (RNFNN) and a modified LPNN estimators for the problem of target localization in distributed MIMO radar. This methodology of RNFNN will also be extended to the scenario of antenna position uncertainties. We first formulate the target localization as an unconstrained ML estimation problem, which has a complicated objective function and is tricky to optimize directly. By introducing some auxiliary variables and taking some manipulations, this unconstrained optimization problem is then transformed into a tractable one with equality constraints. Then, two energy functions are constructed by making use of different strategies, and their corresponding neural networks, i.e. the RNFNN and modified LPNN, are defined and compared. It is attractive that the proposed RNFNN implementation has much simpler structure and faster convergence speed.
Numerical simulations verify the efficiency of the proposed approach by achieving the CRLB for high, moderate, and relatively low SNR levels. Moreover, as demonstrated in Section \ref{sec:simulation} through different simulation scenarios, both the proposed RNFNN and modified LPNN approaches have merits in localization accuracy and computational complexity compared to the existing ones.
The main contributions of this paper are summarized as
follows:
\begin{itemize}
	\item We develop a novel analog neural network approach to solving the problem of target localization in distributed MIMO radar. The convergence of the proposed approach is much faster while delivering enhanced performance.
	\item We present a modified LPNN estimator base on the existing one to deal with the more general noise cases with non-identical noise variances.
	\item Some theoretical analysis is carried out to show that the proposed RNFNN with much simpler structure has faster convergence speed.
	\item The proposed RNFNN approach is extended for the localization problem in the presence of antenna position uncertainties.
	\item Comparisons with state-of-the-art methods are carried out, which demonstrate that the proposed method is very attractive in terms of localization performance and computational complexity.
\end{itemize}

	
The rest of the paper is organized as follows. Section \ref{sec_model} introduces the measurement model of target localization in distributed MIMO radar. In Section \ref{sec_method}, two different energy functions are constructed and the corresponding neural networks for target localization are defined and compared. Next, their computational complexity is analyzed as well. In Section \ref{sec_Extension}, a solution is proposed for target localization with antenna position errors. In Section \ref{sec:simulation}, numerical simulations are conducted to evaluate the performance of the proposed approach and compare it with state-of-the-art methods, followed by some conclusions drawn in Section \ref{sec_conclusion}.

The notations used throughout this paper are as follows. We use bold lower-case, bold upper-case and italic letters to denote the column vector, the matrix, and the scalar, respectively. The $ m $-th entry of $ \boldsymbol{a} $ is denoted by $ \boldsymbol{a}_{m} $ and the $ (m,n) $-th entry of $ \boldsymbol{A} $ is denoted by $ \boldsymbol{A}_{mn} $. The symbol $ (\cdot)^{\text{T}} $  represents the transpose operator. $ \boldsymbol{I}_{M} $ represents an $ M \times M $  identity matrix. $ \boldsymbol{1}_{M \times N} $ and $ \boldsymbol{0}_{M \times N} $ denote respectively an $ M \times N $ matrix whose entries are all one and an $ M \times N $ matrix with all entries being zero. $ \mathbb{R}^{M} $ is the set of $ M $-dimensional real vectors. The notation $ diag(\boldsymbol{a}) $ denotes the diagonal matrix with entries of $ \boldsymbol{a} $ on its diagonal. The letters $ m $ and $ n $ often serve as index. $ \left\|  \boldsymbol{x} \right\|_2 $ and $ \left\|  \boldsymbol{x} \right\|_1 $ are $ \ell_2 $ and $ \ell_1 $ norms of $ \boldsymbol{x} $. lg($ \cdot $) is the logarithm with base 10.
In addition, given a real scalar function $ f(\boldsymbol{x}) $, $ \boldsymbol{x} \in \mathbb{R}^{M} $, $ \nabla f(\boldsymbol{x}) = [\frac{\partial f(\boldsymbol{x})}{\partial x_{1}}, \ldots, \frac{\partial f(\boldsymbol{x})}{\partial x_{M}} ]^{\text{T}} \in \mathbb{R}^{M} $ and $ \nabla_{\boldsymbol{x}\boldsymbol{x}}^2 f(\boldsymbol{x}) = \frac{\partial^2 f(\boldsymbol{x})}{\partial \boldsymbol {x} \partial \boldsymbol{x}^{\text{T}}} = \frac{\partial}{\partial \boldsymbol{x}} [ \frac{\partial f(\boldsymbol{x})}{\partial \boldsymbol{x}^{\text{T}}} ]  \in \mathbb{R}^{M \times M} $ denote respectively the gradient vector and the Hessian matrix with respect to $ \boldsymbol{x} $.


\section{Measurement Model} \label{sec_model}
\begin{figure}[!t]
	\centering
	\includegraphics[width=3.5in]{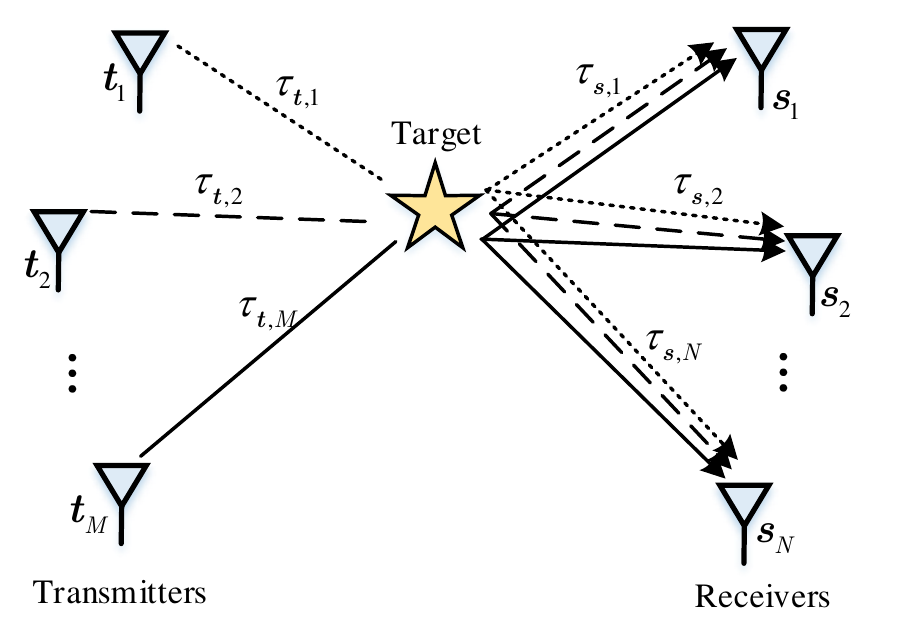}
	\caption{ Illustration of target localization using a distributed MIMO radar consisting of $ M $ transmitters and $ N $ receivers\protect\footnotemark[1]. }
	\label{fig:system_model}
\end{figure}

We consider the localization of a single target using a distributed MIMO radar system consisting of $ M $ transmitters and $ N $ receivers in a $ D $-dimensional space ($ D = 2 $ or $ 3 $). As illustrated in Fig. \ref{fig:system_model}, the actual transmitter and receiver positions are known and denoted by $ \boldsymbol{t}_m  \in \mathbb{R}^{D} $ and  $ \boldsymbol{s}_n  \in \mathbb{R}^{D} $  for $ m=1,\ldots,M $ and $ n=1,\ldots,N $, respectively. The transmitters send out $ M $ narrowband probing signals, which are reflected by a target at unknown location $ \boldsymbol{u}  \in \mathbb{R}^{D} $. The $ N $ receivers sense the reflected signals from the target to determine the location of the target. As a common assumption \cite{2010-He-NoncoherentMIMORadar,2016-Khomchuk-Performanceanalysistarget}, the transmitted signals are approximately orthogonal and maintain approximate orthogonality for the allowed time delays, so that the signals from different transmitters can be separated by matched filtering at each receiver. In addition, the Doppler shift effect can be ignored, which is reasonable for free space radio transmission, where the signal propagates at the speed of light \cite{2019-Jia-EffectSensorMotion}.

\footnotetext[1]{When conducting 3D localization, the transmitters and receivers should be distributed in the 3D space as well.}

Let $ \tau_{\boldsymbol{t},m} $ denote the time that the signal transmitted by the $ m $-th transmitter travels to the target and $ \tau_{\boldsymbol{s},n} $ denotes the time that reflected signal travels from the target to the $ n $-th receiver. Therefore, the noise-free TD for the pair of the $ m $-th transmitter and the $ n $-th receiver is
\begin{equation}
\tau_{mn} = \tau_{\boldsymbol{t},m} + \tau_{\boldsymbol{s},n},\quad m = 1,\ldots,M, n = 1,\ldots,N.
\end{equation}
After multiplying by the known signal propagation speed $ c $, the noise-free bistatic range (BR) is given as bellow
\begin{equation} \label{eq:BR_def}
\begin{aligned}
{r}_{mn} &= c \tau_{mn} \\
& = \left\| \boldsymbol{u} - \boldsymbol{t}_m \right\|_2 + \left\| \boldsymbol{u} - \boldsymbol{s}_n \right\|_2, \\
& \quad m = 1,\ldots,M, n = 1,\ldots,N.
\end{aligned}
\end{equation}

Taking into account measurement errors, the measured version of $ {r}_{mn} $ is modeled as
\begin{equation} \label{eq:BRmesurement_model}
\begin{aligned}
\tilde{r}_{mn} & = {r}_{mn} + \Delta r_{mn}, \\
& = \left\| \boldsymbol{u} - \boldsymbol{t}_m \right\|_2 + \left\| \boldsymbol{u} - \boldsymbol{s}_n \right\|_2  + \Delta r_{mn}, \\
& \quad m = 1,\ldots,M, n = 1,\ldots,N
\end{aligned}
\end{equation}
where $ \Delta \textbf{r} = \left[ \Delta r_{11},\Delta r_{12},..., \Delta r_{NM} \right]^\text{T}  $ represents the BR measurement errors and is modeled as zero-mean Gaussian random vector with covariance matrix $ \textbf{Q}_r = \left[ \sigma_{11}^2,\sigma_{12}^2,..., \sigma_{NM}^2 \right]^\text{T} $. Assume that $ \textbf{Q}_r $ is known or accurately estimated, which is a commonplace assumption in the localization literature \cite{2010-Godrich-TargetLocalizationAccuracy,2012-Niu-TargetLocalizationTracking,2014-Rui-EllipticLocalizationPerformance,2009-Godrich-Targetlocalisationtechniques,2013-Dianat-TargetLocalizationusing,2015-Einemo-Weightedleastsquares,2015-Noroozi-TargetLocalizationBistatic,2017-Amiri-AsymptoticallyEfficientTarget,2020-Noroozi-ClosedFormSolution,2017-Noroozi-IterativeTargetLocalization,2016-Liang-LagrangeProgrammingNeural}. Our aim is to find the target position $ \boldsymbol{u} $ based on the transmitter and receiver positions $ \{ \boldsymbol{t}_m \} $, $ \{ \boldsymbol{s}_n \} $ and BR measurements $ \left\lbrace \tilde{r}_{mn}\right\rbrace $.  It should be noted that due to nonconvex and nonlinear nature of the associated ML estimation problem, its globally optimal solution is difficult to obtain \cite{2015-Noroozi-TargetLocalizationBistatic}.

In the case of multi-target situation, the localization task can be carried out in two steps \cite{2020-Noroozi-EfficientClosedForm} : 1) a data association algorithm (e.g., \cite{2015-Yi-MIMOPassiveRadar,2021-Kazemi-DataAssociationMulti}) is first used to separate the measurements corresponding to each target, and 2) single-target localization is then performed.

\section{Neural Network Approach to Target Localization} \label{sec_method}
Since networks with energy functions intrinsically try to minimize a mathematical function, if one can map the problem to be minimized onto the network in a direct way, then it can find the minimum by obeying its own dynamics \cite{1988-Hopfield-Artificialneuralnetworks}. Therefore, the key step of formulating the localization problem as an optimization neural network is to construct an appropriate energy function such that the lowest energy state corresponds to the true target location. In the following, we first formulate this localization problem as an unconstrained ML estimation problem, which has a complicated objective function and is tricky to optimize directly. By introducing some auxiliary variables and taking some manipulations, this unconstrained optimization problem is then transformed into a more tractable one with equality constraints. Next, two energy functions are constructed by making use of different strategies, and their corresponding neural networks are defined and compared.

\subsection{Energy Function} \label{subsec:EF}
Given BR measurements $ \left\lbrace  \tilde{r}_{mn}\right\rbrace $, $ m = 1,\ldots,M $, $ n = 1,\ldots,N $, the ML estimate of the target location is obtained by minimizing an objective function, i.e.,
\begin{equation}  \label{eq:OPOP}  
\min \limits_{\boldsymbol{u}} \quad \sum_{m=1}^M \sum_{n=1}^N w_{mn} \left(\tilde{r}_{mn}-  \left\| \boldsymbol{u} - \boldsymbol{t}_m \right\|_2 - \left\| \boldsymbol{u} - \boldsymbol{s}_n \right\|_2  \right)^2
\end{equation}
where
\begin{equation}  
w_{mn} = \dfrac{1/\sigma_{mn}^2}{\sum_{m=1}^M \sum_{n=1}^N \{1/\sigma_{mn}^2 \}}
\end{equation}
is the weighting factor (when there is no information of $ \textbf{Q}_r $, set $ w_{mn}=1 $ for all $ m $ and $ n $). The objective function in (\ref{eq:OPOP}) is highly nonconvex and nonlinear. 
Instead of optimizing $ \boldsymbol{u} $ directly, we introduce auxiliary variables $ g_{\boldsymbol{t},m} $ and $ g_{\boldsymbol{s},n} $
\begin{equation} \label{eq:gt}
g_{\boldsymbol{t},m} = \left\| \boldsymbol{u} - \boldsymbol{t}_m \right\|_2, \quad m = 1,\ldots,M,
\end{equation}
\begin{equation} \label{eq:gs}
g_{\boldsymbol{s},n} = \left\| \boldsymbol{u} - \boldsymbol{s}_n \right\|_2, \quad n = 1,\ldots,N,
\end{equation}
and arrange them into column vectors $ \boldsymbol{g_t} = [g_{\boldsymbol{t},1},\ldots, g_{\boldsymbol{t},M}]^{\text{T}} $ and $ \boldsymbol{g_s}= [g_{\boldsymbol{s},1},\ldots, g_{\boldsymbol{s},N}]^{\text{T}} $.
By making use of (\ref{eq:gt}) and (\ref{eq:gs}), the unconstrained optimization problem (\ref{eq:OPOP}) can be reformulated as the following constrained optimization problem
\begin{equation} \label{eq:COP1} 
\begin{aligned}
\min \limits_{\boldsymbol{u},\boldsymbol{g_t},\boldsymbol{g_s}} \quad  &  \sum_{m=1}^M \sum_{n=1}^N w_{mn} \left(\tilde{r}_{mn}
- g_{\boldsymbol{t},m} - g_{\boldsymbol{s},n} \right)^2 \\
s.t. \quad  g_{\boldsymbol{t},m} = & \left\| \boldsymbol{u} - \boldsymbol{t}_m \right\|_2, \quad m = 1,\ldots,M,\\
\quad g_{\boldsymbol{s},n} = & \left\| \boldsymbol{u} - \boldsymbol{s}_n \right\|_2, \quad n = 1,\ldots,N. 
\end{aligned}
\end{equation}

Note that the gradient vectors of terms $ \left\| \boldsymbol{u} - \boldsymbol{t}_m \right\|_2 $ and $ \left\| \boldsymbol{u} - \boldsymbol{s}_n \right\|_2 $ contain some factors that are in terms of $ \left( 1/ \left\| \boldsymbol{u} - \boldsymbol{t}_m \right\|_2 \right) $ and $ \left( 1/\left\| \boldsymbol{u} - \boldsymbol{s}_n \right\|_2 \right) $. Hence, when the current estimate of $ \boldsymbol{u} $ is close to either $ \boldsymbol{t}_m $ or $ \boldsymbol{s}_n $, we have the ill-posed problem, since some elements of the gradient vector may approach infinity. To remove the ill-posed situation, we take the square operation on both sides of the equality constraints. The problem (\ref{eq:COP1}) is then transformed into
\begin{equation} \label{eq:COP2} 
	\begin{aligned}
	\min \limits_{\boldsymbol{u},\boldsymbol{g_t},\boldsymbol{g_s}} \quad  & \sum_{m=1}^M \sum_{n=1}^N w_{mn} \left(\tilde{r}_{mn}
	- g_{\boldsymbol{t},m} - g_{\boldsymbol{s},n} \right)^2 \\
	s.t. \quad g_{\boldsymbol{t},m}^2 = & \left\| \boldsymbol{u} - \boldsymbol{t}_m \right\|_2^2, \quad m = 1,\ldots,M,\\
	g_{\boldsymbol{s},n}^2 = & \left\| \boldsymbol{u} - \boldsymbol{s}_n \right\|_2^2, \quad n = 1,\ldots,N, \\
	g_{\boldsymbol{t},m} \geq & 0, \quad m = 1,\ldots,M, \\
	g_{\boldsymbol{s},n} \geq & 0, \quad n = 1,\ldots,N.  
	\end{aligned}
\end{equation}


Note that the auxiliary variables $ g_{\boldsymbol{t},m} $ and $ g_{\boldsymbol{s},n} $ respectively correspond to the distances between the target and the $ m $-th transmitter, and the distance between the
target and the $ n $-th receiver. Observing the objective function and employing the measurement property $ \tilde{r}_{mn} \geq 0 $, the inequality constraints in (9) can be removed [21]. As a result, (\ref{eq:COP2}) is equivalent to
\begin{equation} \label{eq:COP3}
	\begin{aligned}
	\min \limits_{\boldsymbol{u},\boldsymbol{g_t},\boldsymbol{g_s}} \quad & \sum_{m=1}^M \sum_{n=1}^N w_{mn} \left(\tilde{r}_{mn}
	- g_{\boldsymbol{t},m} - g_{\boldsymbol{s},n} \right)^2 \\
	s.t. \quad g_{\boldsymbol{t},m}^2 = & \left\| \boldsymbol{u} - \boldsymbol{t}_m \right\|_2^2, \quad m = 1,\ldots,M,\\
	g_{\boldsymbol{s},n}^2 = & \left\| \boldsymbol{u} - \boldsymbol{s}_n \right\|_2^2, \quad n = 1,\ldots,N. 
	\end{aligned}
\end{equation}

To solve the above constrained	optimization problem, a  Lagrangian function can be constructed, i.e.,
\begin{equation} \label{eq:LPF}
\begin{aligned}
L_w(\boldsymbol{y},\boldsymbol{\lambda}) = &\, \dfrac{1}{2} \sum_{m=1}^M \sum_{n=1}^N w_{mn} \left( \tilde{r}_{mn} - g_{\boldsymbol{t},m} - g_{\boldsymbol{s},n} \right) ^2 \\
& + \sum_{m=1}^M \lambda_{\boldsymbol{t},m} \left( g_{\boldsymbol{t},m}^2 -\left\| \boldsymbol{u} - \boldsymbol{t}_m \right\|_2^2 \right) \\
& + \sum_{n=1}^N \lambda_{\boldsymbol{s},n} \left(  g_{\boldsymbol{s},n}^2 -\left\| \boldsymbol{u} - \boldsymbol{s}_n \right\|_2^2 \right) \\
\end{aligned}
\end{equation}
where $ \boldsymbol{y} = \left[  \boldsymbol{u}^{\text{T}}, \boldsymbol{g}^{\text{T}}_{\boldsymbol{t}}, \boldsymbol{g}^{\text{T}}_{\boldsymbol{s}} \right]^{\text{T}} $ is the variable vector, and $\boldsymbol{\lambda} = [ \boldsymbol{\lambda}^{\text{T}}_{\boldsymbol{t}}, \boldsymbol{\lambda}^{\text{T}}_{\boldsymbol{s}} ]^{\text{T}} $ with $ {\boldsymbol{\lambda_s}} = [\lambda_{\boldsymbol{t},1},\ldots, \lambda_{\boldsymbol{t},M}]^{\text{T}} \in \mathbb{R}^{M} $ and $ \boldsymbol{\lambda_r}= [\lambda_{\boldsymbol{s},1},\ldots, \lambda_{\boldsymbol{s},N}]^{\text{T}} \in \mathbb{R}^{N} $ is Lagrange multiplier vector. As the Lagrange multiplier terms in (\ref{eq:LPF}) are nonconvex (which is demonstrated in Section \ref{subsec:Comparison} ), inspired by \cite{1992-Zhang-Lagrangeprogrammingneural,2018-Han-AugmentedLagrangeProgramming}, we can improve the convexity around the minimum point of the problem through introducing an augmented term 
\begin{equation} \label{eq:augmented_term}
\begin{aligned}
A(\boldsymbol{y},\boldsymbol{\lambda}) = & \dfrac{C}{4}\sum_{m=1}^M \left( g_{\boldsymbol{t},m}^2 - \left\| \boldsymbol{u} - \boldsymbol{t}_m \right\|_2^2 \right) ^2 \\
& + \dfrac{C}{4} \sum_{n=1}^N \left(  g_{\boldsymbol{s},n}^2 -\left\| \boldsymbol{u} - \boldsymbol{s}_n \right\|_2^2 \right) ^2
\end{aligned}
\end{equation}
into the Lagrangian function, where $ C $ is a positive constant that controls the amplitude of the augmented term. The Lagrangian function is now modified as
\begin{equation} \label{eq:ALPF}
 L_{Aw}(\boldsymbol{y},\boldsymbol{\lambda}) = L_w(\boldsymbol{y},\boldsymbol{\lambda}) + A(\boldsymbol{y},\boldsymbol{\lambda}).
\end{equation}

It is noted that, compared to the LPNN method in \cite{2016-Liang-LagrangeProgrammingNeural} and \cite{2020-Shi-RobustMIMORadar}, the Lagrangian function given in this paper is weighted by $ w_{mn} $, and the relevant method is called modified LPNN, which is useful to deal with the more general noise cases with non-identical noise variances and improves significantly the positioning accuracy. This will be demonstrated in Section \ref{sec:simulation}.

However, the convergence of the method with Lagrangian objective function is relatively slow, as the update of the Lagrangian multiplier is an ascent iteration and it can only converge moderately quickly \cite{2013-Zhang-Matrixanalysisapplications}. Having a close look at \eqref{eq:COP3} and recall the measurement model (\ref{eq:BRmesurement_model}), one can find that the exact equality constraints are not necessary due to measurement errors. In the following, we relax the problem (\ref{eq:COP3}) by defining variables $ {\boldsymbol{h_t}} = [h_{\boldsymbol{t},1},\ldots, h_{\boldsymbol{t},M}]^{\text{T}} $ and $ \boldsymbol{h_s}= [h_{\boldsymbol{s},1},\ldots, h_{\boldsymbol{s},N}]^{\text{T}} $. We require that $h^2_{\boldsymbol{t},m}$ is close to $g^2_{\boldsymbol{t},m}$ as much as possible and $h^2_{\boldsymbol{s},n}$ is close to $g^2_{\boldsymbol{s},n}$ as much as possible, while minimizing
\begin{equation}
\sum_{m=1}^M \sum_{n=1}^N w_{mn} \left( \tilde{r}_{mn} - h_{\boldsymbol{t},m} - h_{\boldsymbol{s},n} \right) ^2 
\end{equation}
which leads to the minimization of the following relaxed energy function (RNF) :
\begin{equation} \label{eq:RNF}
\begin{aligned}
E(\boldsymbol{x}) = & \dfrac{1}{2} \sum_{m=1}^M \sum_{n=1}^N w_{mn} \left( \tilde{r}_{mn} - h_{\boldsymbol{t},m} - h_{\boldsymbol{s},n} \right) ^2 \\
& + \dfrac{\rho}{4} \sum_{m=1}^M \left( h_{\boldsymbol{t},m}^2 -\left\| \boldsymbol{u} - \boldsymbol{t}_m \right\|_2^2 \right)  ^2 \\
& + \dfrac{\rho}{4} \sum_{n=1}^N \left(  h_{\boldsymbol{s},n}^2 -\left\| \boldsymbol{u} - \boldsymbol{s}_n \right\|_2^2 \right)  ^2 \\
\end{aligned}
\end{equation}
where $ \boldsymbol{x} = \left[  \boldsymbol{u}^{\text{T}},  \boldsymbol{h}^{\text{T}}_{\boldsymbol{t}}, \boldsymbol{h}^{\text{T}}_{\boldsymbol{s}} \right] ^{\text{T}} \in \mathbb{R}^{D+M+N} $, $ \rho > 0 $ is a constant penalty parameter, and the coefficients 1/2 and 1/4 are used to simplify some expressions later. 


\begin{figure*}[!t]
	\centering
	\includegraphics[width=6.5in]{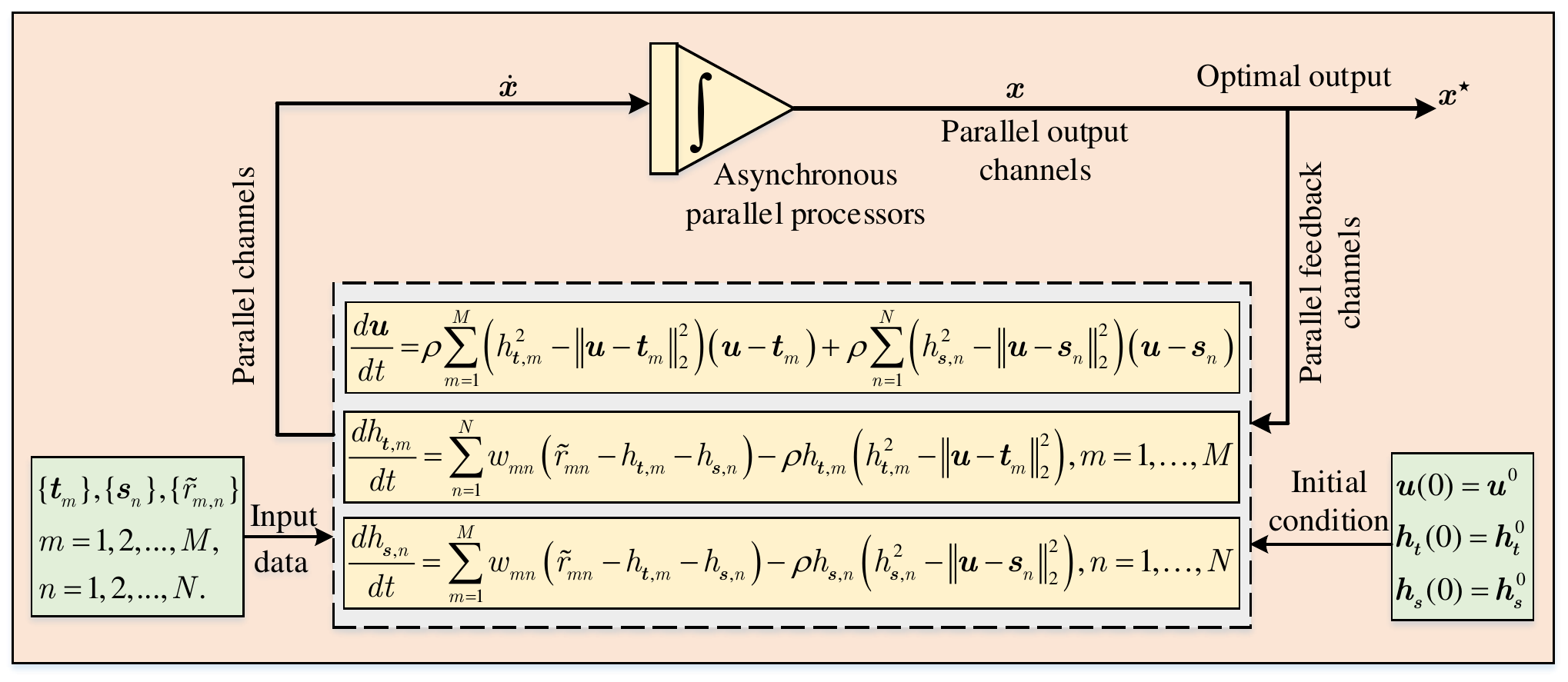}
	\caption{Diagram of the relaxed energy function based neural network approach to target localization in distributed MIMO radar.}
	\label{fig:diagram_RNFNN}
\end{figure*}

\subsection{Definition of the Network} \label{subsec:Def_Network}
Our aim now is to design an analog neural network that will settle down to an equilibrium point satisfying the first-order necessary condition of optimality. Inspired by \cite{1988-Hopfield-Artificialneuralnetworks,1986-Tank-SimpleNeuralOptimization,2003-Leung-highperformancefeedback}, the transient behavior of the neural network is defined by the following equation.
\begin{equation} \label{eq:def_RNFNN}
\dfrac{d\boldsymbol{x}}{dt} = - \nabla E(\boldsymbol{x}).
\end{equation}
where $ t $ is the time variable.

If the network is physically stable, the equilibrium point $ \boldsymbol{x}^\star = [ \boldsymbol{u}^\star; \boldsymbol{h}^{\star}_{\boldsymbol{t}}; \boldsymbol{h}^{\star}_{\boldsymbol{s}} ] $, described by $ \left.  (d\boldsymbol{x}/dt)\right| _{\boldsymbol{x}^\star} = \boldsymbol{0} $, obviously meets the necessary condition and thus provides a solution to the original positioning problem.

Expressing the definition (\ref{eq:def_RNFNN}) in a component form, we obtain the RNFNN described as follows.

\textit{State equations of RNFNN:}
\begin{equation} \label{eq:stateEq_RNFNN}
\left\{
\begin{aligned}
\dfrac{d\boldsymbol{u}}{dt} = \,& \rho \sum_{m=1}^M \left( h_{\boldsymbol{t},m}^2 -\left\| \boldsymbol{u} - \boldsymbol{t}_m \right\|_2^2 \right) \left( \boldsymbol{u} - \boldsymbol{t}_m \right) \\
 & + \rho \sum_{n=1}^N \left( h_{\boldsymbol{s},n}^2 -\left\| \boldsymbol{u} - \boldsymbol{s}_n \right\|_2^2 \right) \left( \boldsymbol{u} - \boldsymbol{s}_n\right), \\
\dfrac{d h_{\boldsymbol{t},m} }{dt} = & \sum_{n=1}^N w_{mn} \left( \tilde{r}_{mn} - h_{\boldsymbol{t},m} - h_{\boldsymbol{s},n} \right) \\
& -\rho h_{\boldsymbol{t},m} \left( h_{\boldsymbol{t},m}^2 -\left\| \boldsymbol{u} - \boldsymbol{t}_m \right\|_2^2 \right), \\
&  \quad m = 1,\ldots,M, \\
\dfrac{d h_{\boldsymbol{s},n} }{dt} = & \sum_{m=1}^M w_{mn} \left( \tilde{r}_{mn} - h_{\boldsymbol{t},m} - h_{\boldsymbol{s},n} \right) \\
& -\rho h_{\boldsymbol{s},n} \left( h_{\boldsymbol{s},n}^2 -\left\| \boldsymbol{u} - \boldsymbol{s}_n \right\|_2^2 \right), \\
&  \quad n = 1,\ldots,N.
\end{aligned}	
\right.
\end{equation}
where variables $ \boldsymbol{u} = \{x,y\} $ for $ D=2 $ or $ \boldsymbol{u} = \{x,y,z\} $ for $ D=3 $, and $ \boldsymbol{h_t} = \{ h_{\boldsymbol{t},m} \} $, $  m = 1,\ldots,M $, $ \boldsymbol{h_s} = \{ h_{\boldsymbol{s},n} \} $,  $ n = 1,\ldots,N $ are now assigned a physical meaning as neuron activities \cite{1992-Zhang-Lagrangeprogrammingneural}. We collectively refer to them as \textit{variable neurons} $ \boldsymbol{x} = \{ \boldsymbol{u}, \boldsymbol{h_t}, \boldsymbol{h_s} \} $. In the RNFNN, there are $ (D\!+\!M\!+\!N) $ variable neurons, and all neuron dynamics, which can be seen as the state transition of neurons, proceed simultaneously. Fig. \ref{fig:diagram_RNFNN} shows the diagram of the RNFNN approach to target localization in distributed MIMO radar, in which $ \boldsymbol{x}^0 = \left\lbrace \boldsymbol{u}^0, \boldsymbol{h}^0_{\boldsymbol{t}}, \boldsymbol{h}^0_{\boldsymbol{s}} \right\rbrace  $
is the initial condition, and $ \{ \boldsymbol{t}_m \} $, $ \{ \boldsymbol{s}_n \} $, $ \{ \tilde{r}_{mn} \} $, $ m = 1,\ldots,M $, $ n = 1,\ldots,N $ are the input data. The value of parameter $ \rho $ determines the strength of penalty when constraints are violated, which can pull the trajectory of the network into feasible region \cite{1986-Tank-SimpleNeuralOptimization}. A computation is begun by starting at a particular state in the system. That initial state is determined by the program and the input data. The network moves along, following its intrinsic trajectory in its state space. When it gets to the answer (the stable point of the network), it stops, and the answer is read out \cite{1988-Hopfield-Artificialneuralnetworks}.

Different from strictly two-state neuron \cite{1984HopfieldNeuronsgradedresponse}, this network is an analog network with continuous neuron variables and responses. Furthermore, it is a feedback neural network, i.e., the output of \textit{asynchronous} parallel processors return to become the input. It is indicated in \cite{1985-Hopfield-NeuralComputationDecisions} that the effectiveness of neural computing involves both the nonlinear analog responses of neurons and the large degree of connectivity between them. There are three major forms of parallel organization in the system: parallel input (/feedback) channels, parallel output channels, and asynchronous parallel processing elements (integrators).

In a similar manner, using (\ref{eq:ALPF}), the transient behavior of the LPNN is defined by the following equation \cite{1992-Zhang-Lagrangeprogrammingneural}

\begin{equation} \label{eq:def_LPNN1}
\dfrac{d\boldsymbol{y}}{dt} = - \nabla_y L_{Aw}(\boldsymbol{y},\boldsymbol{\lambda}), 
\end{equation}
\begin{equation} \label{eq:def_LPNN2}
\dfrac{d\boldsymbol{\lambda}}{dt} = \nabla_\lambda L_{Aw}(\boldsymbol{y},\boldsymbol{\lambda}).
\end{equation}
Expressing the definition (\ref{eq:def_LPNN1}) and (\ref{eq:def_LPNN2}) in a component form, we obtain the LPNN described as follows.

\textit{State equations of LPNN:}
\begin{equation} \label{eq:stateEq_LPNN}
\left\{
\begin{aligned}
\dfrac{d\boldsymbol{u}}{dt} = & \sum_{m=1}^M \left( 2\lambda_{\boldsymbol{t},m} + C\left( g_{\boldsymbol{t},m}^2 -\left\| \boldsymbol{u} - \boldsymbol{t}_m \right\|_2^2 \right)\right)  \left( \boldsymbol{u} - \boldsymbol{t}_m \right) \\
& + \sum_{n=1}^N \left( 2\lambda_{\boldsymbol{s},n} + C\left(  g_{\boldsymbol{s},n}^2 -\left\| \boldsymbol{u} - \boldsymbol{s}_n \right\|_2^2 \right)\right)  \left( \boldsymbol{u} - \boldsymbol{s}_n\right), \\
\dfrac{d g_{\boldsymbol{t},m} }{dt} & = \sum_{n=1}^N w_{mn} \left( \tilde{r}_{mn} - g_{\boldsymbol{t},m} - g_{\boldsymbol{s},n} \right) - 2g_{\boldsymbol{t},m}\lambda_{\boldsymbol{t},m} \\
& -C g_{\boldsymbol{t},m} \left( g_{\boldsymbol{t},m}^2 -\left\| \boldsymbol{u} - \boldsymbol{t}_m \right\|_2^2 \right), \\
&  \quad m = 1,\ldots,M, \\
\dfrac{d g_{\boldsymbol{s},n} }{dt} & = \sum_{m=1}^M w_{mn} \left( \tilde{r}_{mn} - g_{\boldsymbol{t},m} - g_{\boldsymbol{s},n} \right) - 2g_{\boldsymbol{s},n}\lambda_{\boldsymbol{s},n}\\
& -C g_{\boldsymbol{s},n} \left( g_{\boldsymbol{s},n}^2 -\left\| \boldsymbol{u} - \boldsymbol{s}_n \right\|_2^2 \right), \\
&  \quad n = 1,\ldots,N, \\
\dfrac{d \lambda_{\boldsymbol{t},m} }{dt} & = g_{\boldsymbol{t},m}^2 -\left\| \boldsymbol{u} - \boldsymbol{t}_m \right\|_2^2, \quad m = 1,\ldots,M, \\
\dfrac{d \lambda_{\boldsymbol{s},n} }{dt} & = g_{\boldsymbol{s},n}^2 -\left\| \boldsymbol{u} - \boldsymbol{s}_n \right\|_2^2, \quad n = 1,\ldots,N.
\end{aligned}	
\right.
\end{equation}

Note that, there are extra $ (M\!+\!N) $ \textit{Lagrangian neurons} $\boldsymbol{\lambda} = \{ \boldsymbol{\lambda}_{\boldsymbol{t}}, \boldsymbol{\lambda}_{\boldsymbol{s}} \} $ involved additional to $ (D\!+\!M\!+\!N) $ variable neurons $ \boldsymbol{y} = \{ \boldsymbol{u}, \boldsymbol{g_t}, \boldsymbol{g_s} \} $ in LPNN, which means higher neural circuit complexity or higher computational load demanded \cite{1986-Tank-SimpleNeuralOptimization,2003-Leung-highperformancefeedback}.

\subsection{Comparison of the RNFNN and Modified LPNN} \label{subsec:Comparison}
This subsection provides some theoretical analysis to investigate the properties of the two neural networks defined in Section \ref{subsec:Def_Network}, including the energy function, approximation error and the dynamic behavior of the neural networks. As described in \ref{subsec:EF}, by introducing the variables $ {\boldsymbol{h_t}} $ and $ {\boldsymbol{h_s}} $, the RNFNN avoids the introduction of Lagrange multipliers and multiplier terms. As aforementioned in \ref{subsec:EF}, due to the update of the Lagrange multiplier is an ascend iteration, it can only converge moderately quickly. Next, we will further analyze the Lagrangian multiplier terms, i.e.,
\begin{equation}
L_{\boldsymbol{\lambda_t}}(\boldsymbol{y}) \triangleq \sum_{m=1}^M \lambda_{\boldsymbol{t},m} \left( g_{\boldsymbol{t},m}^2 -\left\| \boldsymbol{u} - \boldsymbol{t}_m \right\|_2^2 \right) 
\end{equation}
and
\begin{equation}
L_{\boldsymbol{\lambda_s}}(\boldsymbol{y}) \triangleq \sum_{n=1}^N \lambda_{\boldsymbol{s},n} \left(  g_{\boldsymbol{s},n}^2 -\left\| \boldsymbol{u} - \boldsymbol{s}_n \right\|_2^2 \right).
\end{equation}
Their Hessian matrices are given as follows.
\begin{equation}
\nabla_{\boldsymbol{yy}}^2 L_{\boldsymbol{\lambda_t}}(\boldsymbol{y}) =
\begin{bmatrix}
-2 \sum_{m=1}^M \lambda_{\boldsymbol{t},m} \cdot \boldsymbol{I}_D & \boldsymbol{0}_{D \times M} \\
\boldsymbol{0}_{M \times D} & 2 diag(\lambda_{\boldsymbol{t},1}, \ldots, \lambda_{\boldsymbol{t},M}) \\
\end{bmatrix}
\end{equation}
and
\begin{equation}
\nabla_{\boldsymbol{yy}}^2 L_{\boldsymbol{\lambda_s}}(\boldsymbol{y}) =
\begin{bmatrix}
-2 \sum_{n=1}^N \lambda_{\boldsymbol{s},n} \cdot \boldsymbol{I}_D & \boldsymbol{0}_{D \times N} \\
\boldsymbol{0}_{N \times D} & 2 diag(\lambda_{\boldsymbol{s},1}, \ldots, \lambda_{\boldsymbol{s},N}) \\
\end{bmatrix}
\end{equation}
We can see that the Hessian matrices $ \nabla_{\boldsymbol{yy}}^2 L_{\boldsymbol{\lambda_t}}(\boldsymbol{y}) $ and $ \nabla_{\boldsymbol{yy}}^2 L_{\boldsymbol{\lambda_s}}(\boldsymbol{y}) $ are not positive definite (or semi-definite), and thus the two Lagrange multiplier terms are non-convex, which may weaken the convexity of $ L_{Aw}(\boldsymbol{y},\boldsymbol{\lambda}) $ and slow down the convergence speed of its corresponding network. In addition, the parameter $ C $ in the augmented term (\ref{eq:augmented_term}) must be large enough to ensure the convexity and stability of $ L_{Aw}(\boldsymbol{y},\boldsymbol{\lambda}) $ at equilibrium point \cite{2016-Liang-LagrangeProgrammingNeural}. However, very large $ C $ would also give rise to the issue of divergence \cite{1992-Zhang-Lagrangeprogrammingneural}.

\begin{figure}[!t]
	\centering
	\subfigure[$ \rho=0.1 $]{
		\includegraphics[width=1.625in]{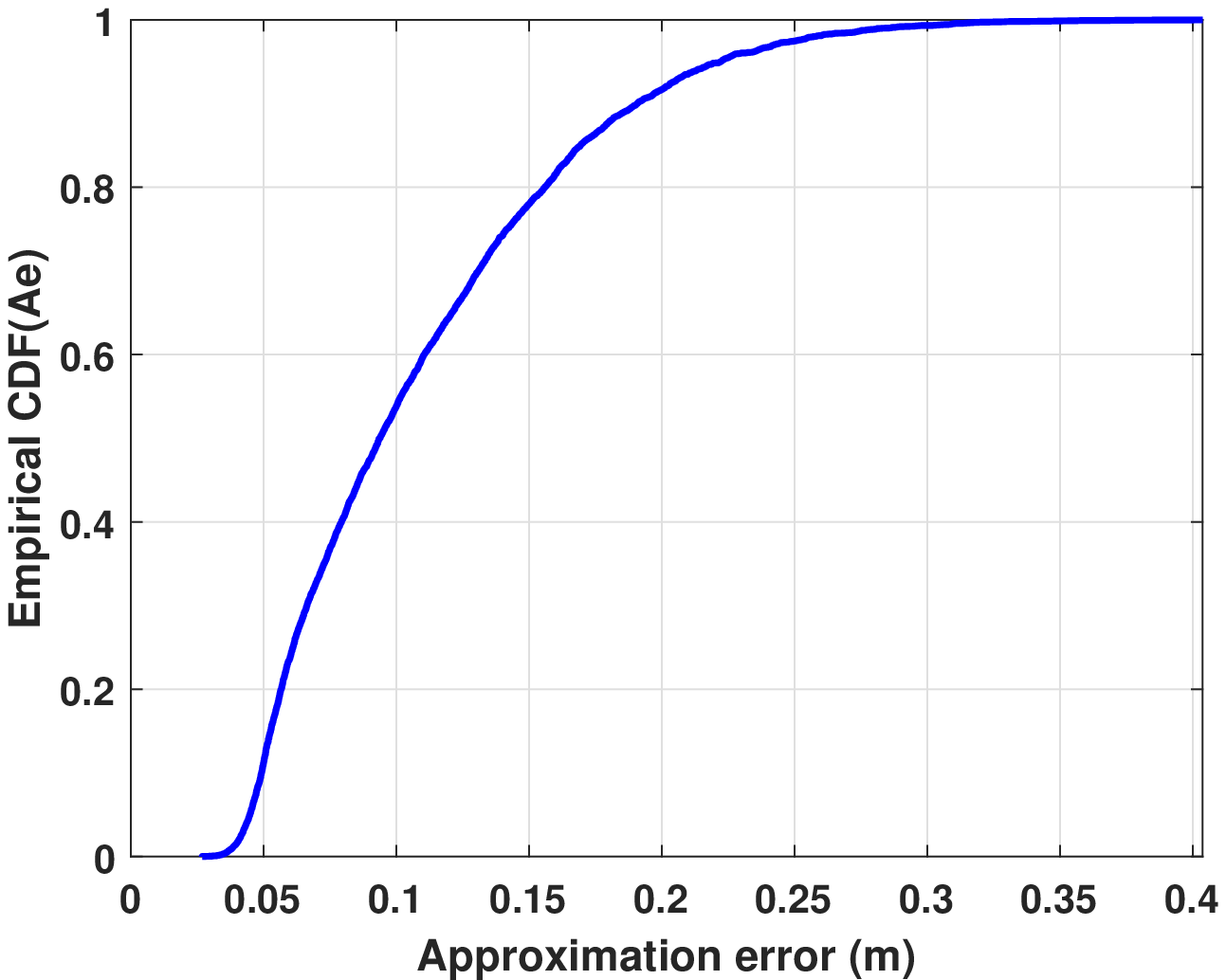}
	}
	\subfigure[$ \rho=0.2 $]{
		\includegraphics[width=1.625in]{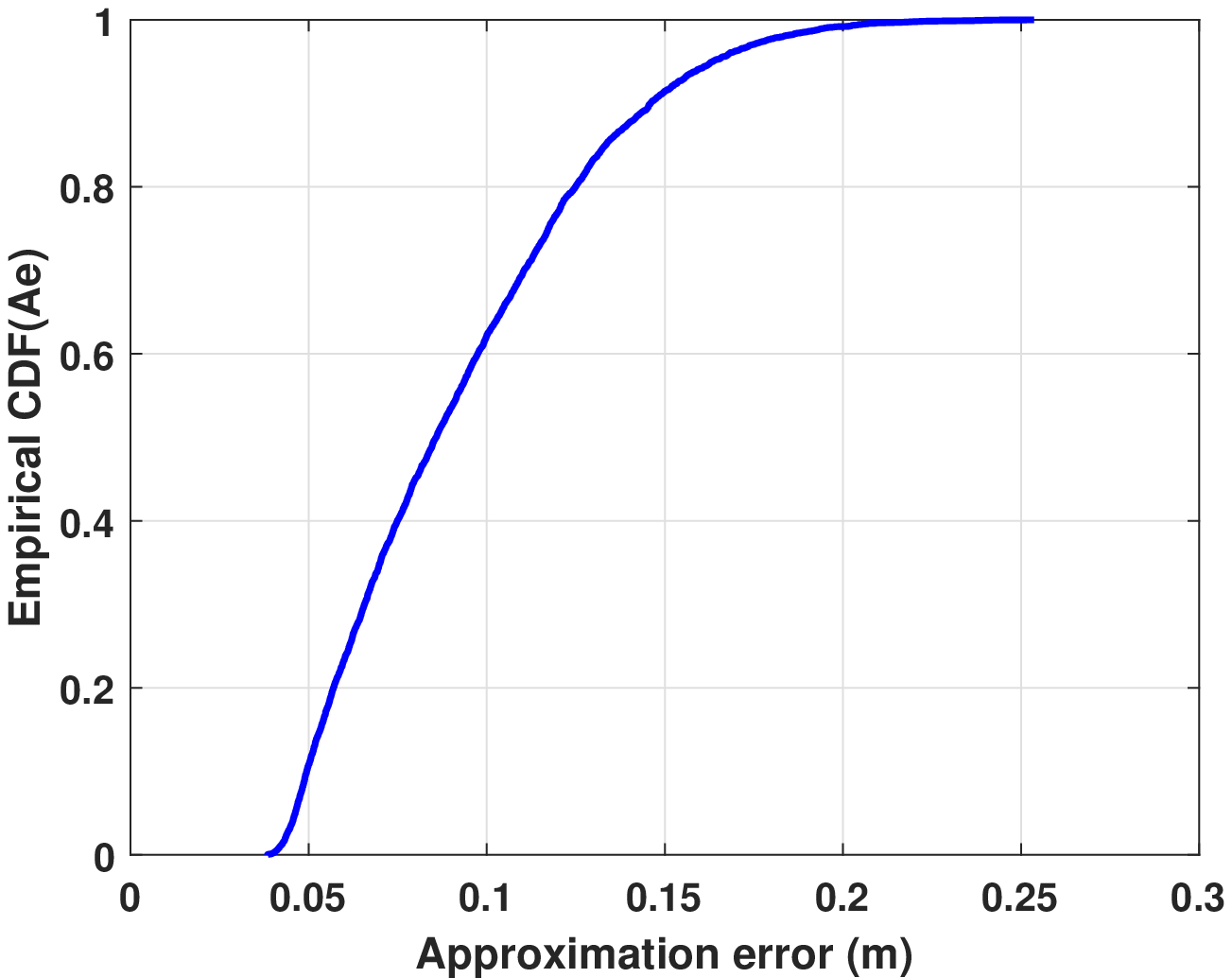}
	}
	\caption{ Empirical cumulative distribution function (CDF) of approximation error when SNR=30dB. }
	\label{fig:Ae}
\end{figure}

Comparing the original optimization problem (\ref{eq:OPOP}) and RNF (\ref{eq:RNF}), we can see that, when $ \rho \to \infty $, the minimization of RNF is equivalent to the original problem (\ref{eq:OPOP}); when $ \rho $ is a finite value, they are not equivalent. However, a finite value of $ \rho $ can bring benefits such as faster convergence while having small approximation errors as shown in Fig. \ref{fig:Ae} (the larger the $ \rho $, the smaller the error, but the more iterations are required). In Fig. \ref{fig:Ae}, the approximation error (Ae) is evaluated through the empirical cumulative distribution function (CDF), which is defined as
\begin{equation}
	\text{CDF}(\text{Ae}) = P \left( \left\|  \hat{\boldsymbol{u}}_{\text{RNFNN}}^{(l)} - \hat{\boldsymbol{u}}_{\text{mLPNN}}^{(l)} \right\|_2 \leq \text{Ae} \right) 
\end{equation}
where $ \hat{\boldsymbol{u}}_{\text{RNFNN}}^{(l)} $ and $ \hat{\boldsymbol{u}}_{\text{mLPNN}}^{(l)} $ denote the target location estimate of the RNFNN and modified LPNN, respectively, for the $ l $-th trial and the number of Monte Carlo trials is $ L = 5000 $.
	
In addition, we note that the root mean square error (RMSE) used in simulations in Section \ref{sec:simulation} is calculated relative to the true target position. The performance of the proposed method is not sensitive to the value of $ \rho $, and we chose $ \rho=0.1 $ in simulations. As shown in Fig. \ref{fig:3D_rmse}, the difference between the RMSE obtained by the RNFNN method and that obtained by the modified LPNN method is very small.

\begin{figure}[!t]
	\centering
	\subfigure[]{
		\includegraphics[width=1.625in]{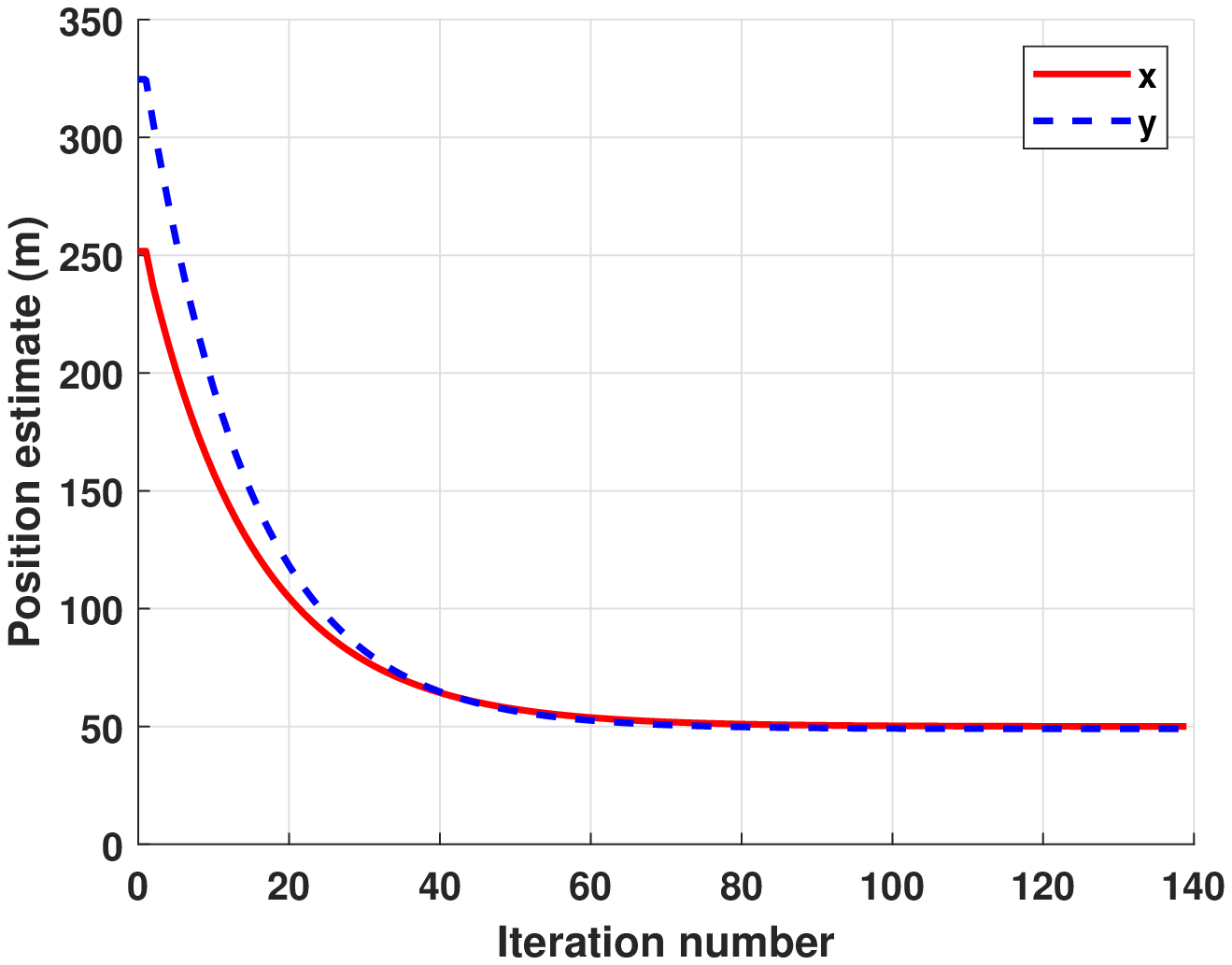}
	}
	\subfigure[]{
		\includegraphics[width=1.625in]{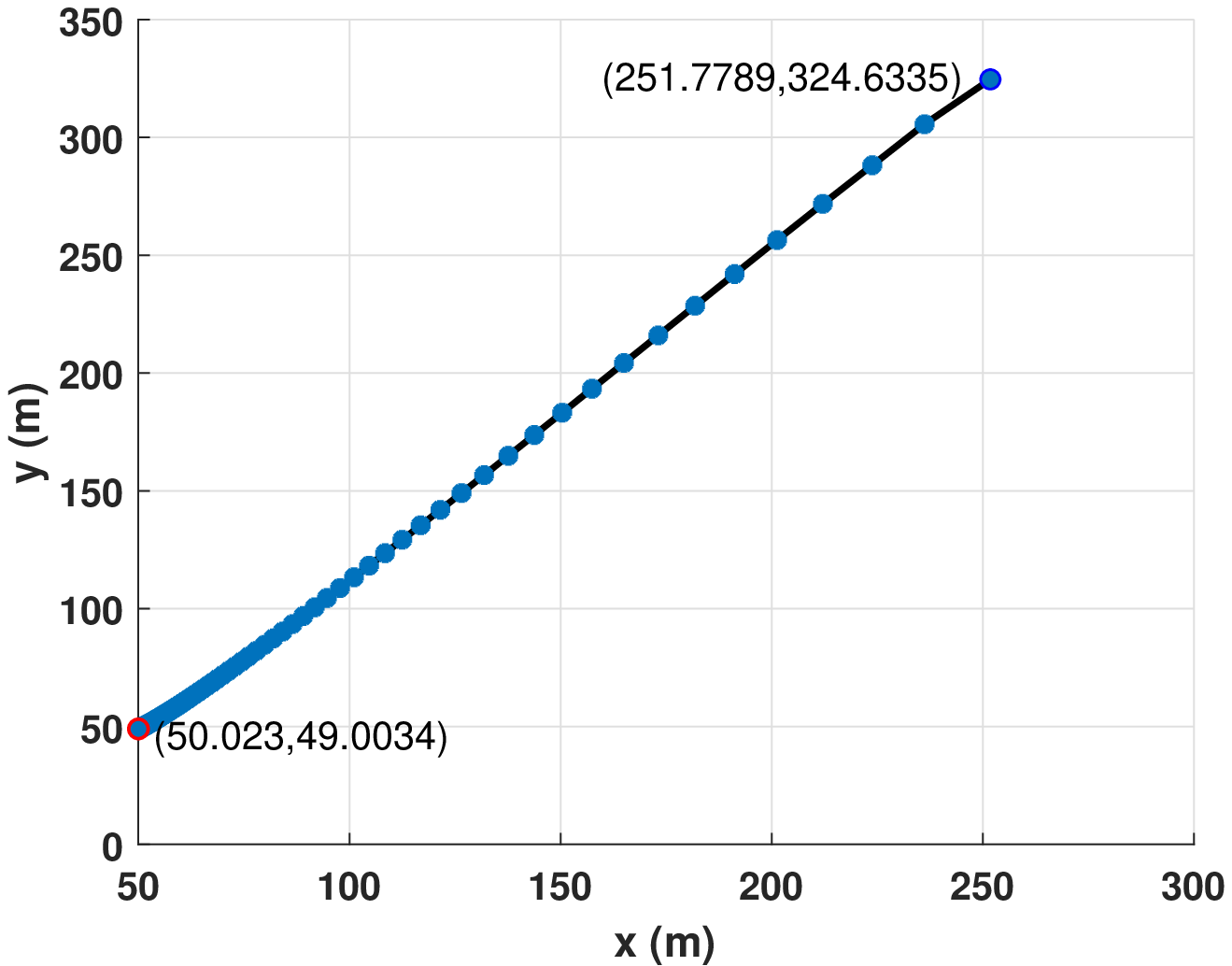}
	}
	\caption{Dynamics and trajectory of estimated $ \boldsymbol{u} = [x, y]^\text{T} $ of RNFNN approach when SNR = 20 dB, starting with a randomly initial value. The true target position is $ [50, 50]^\text{T} $. (a) Dynamics; (b) Trajectory.}
	\label{fig:RNFNN_dynamics}
\end{figure}

\begin{figure}[!t]
	\centering
	\subfigure[]{
		\includegraphics[width=1.625in]{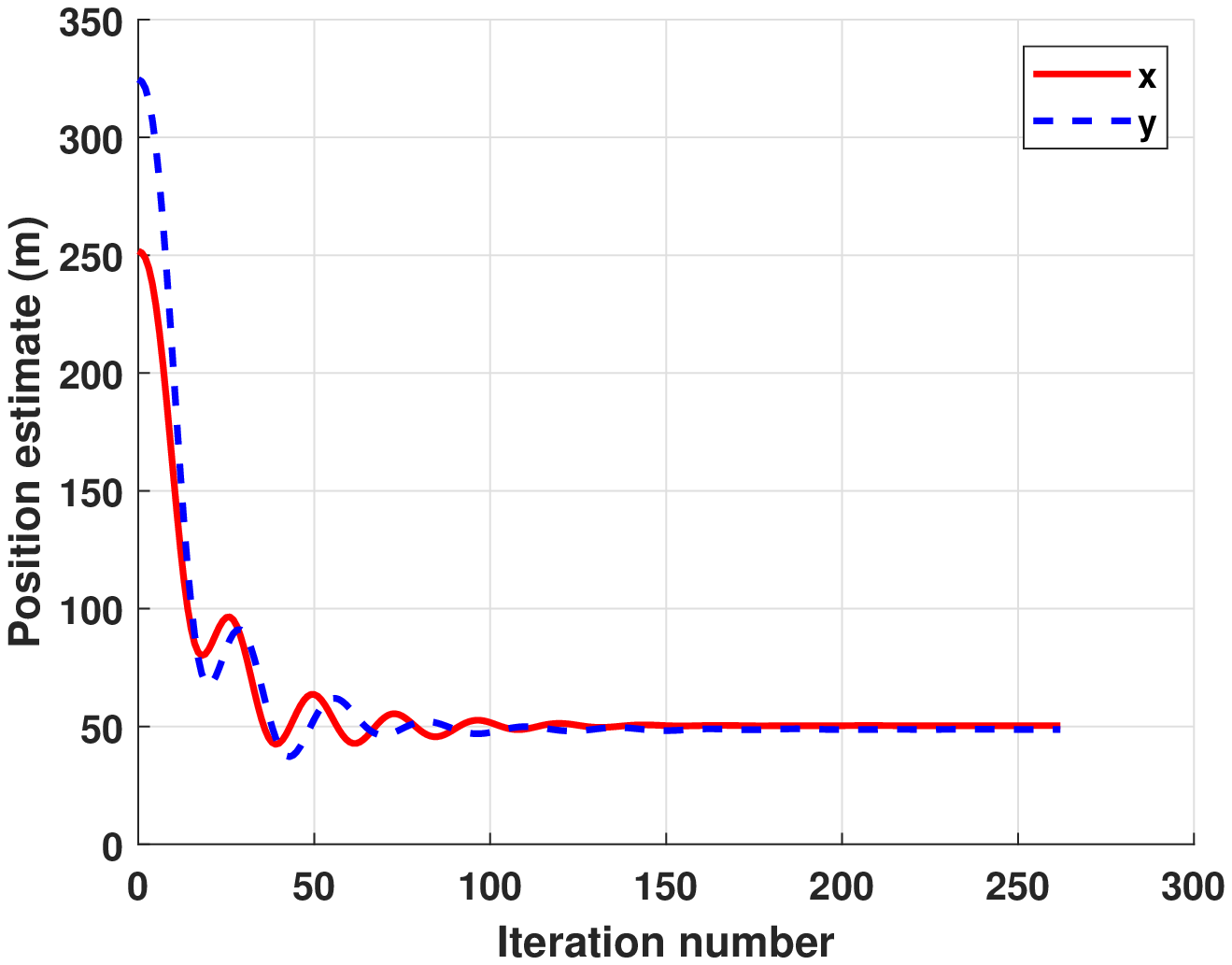}
	}
	\subfigure[]{
		\includegraphics[width=1.625in]{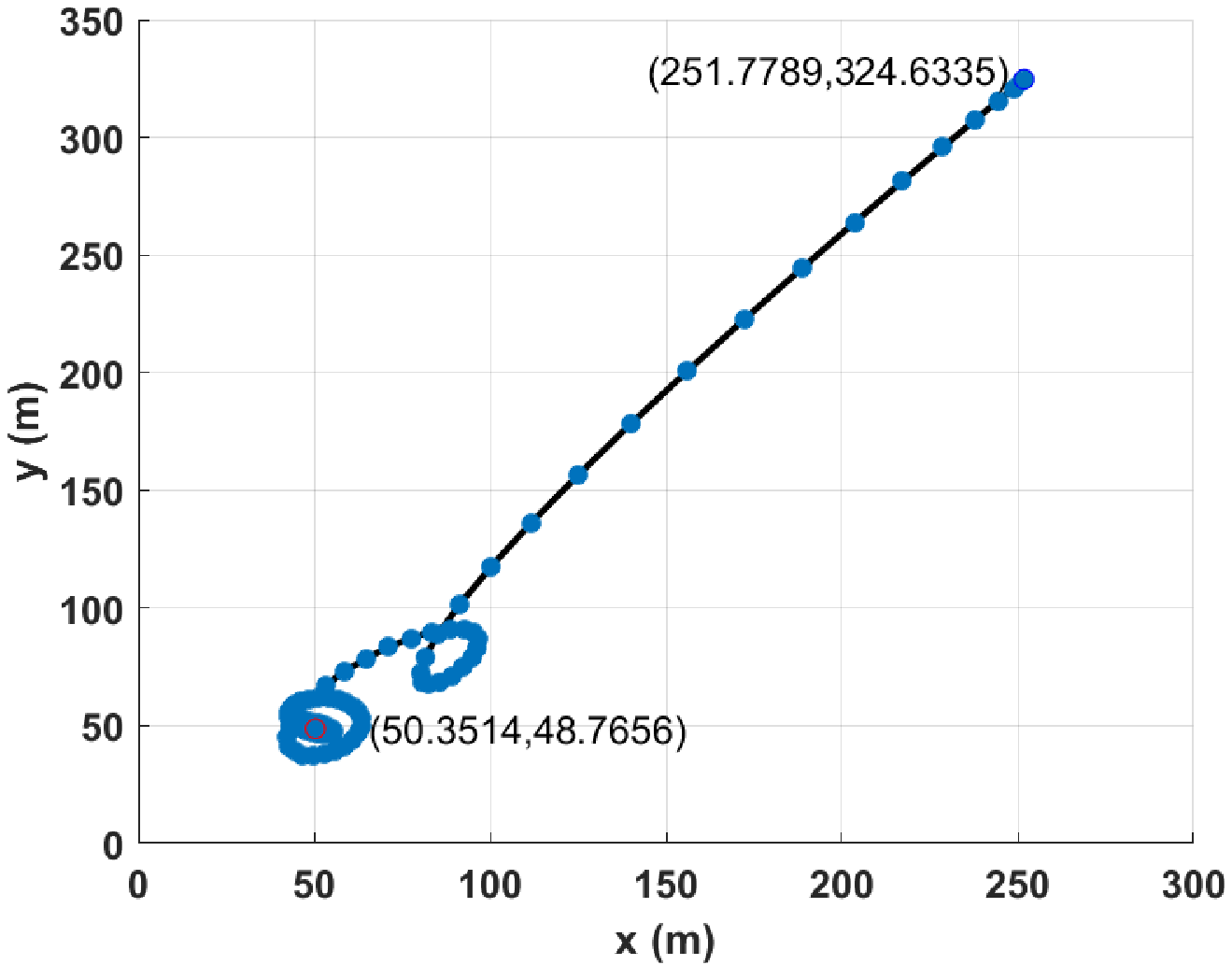}
	}
	\subfigure[]{
		\includegraphics[width=1.625in]{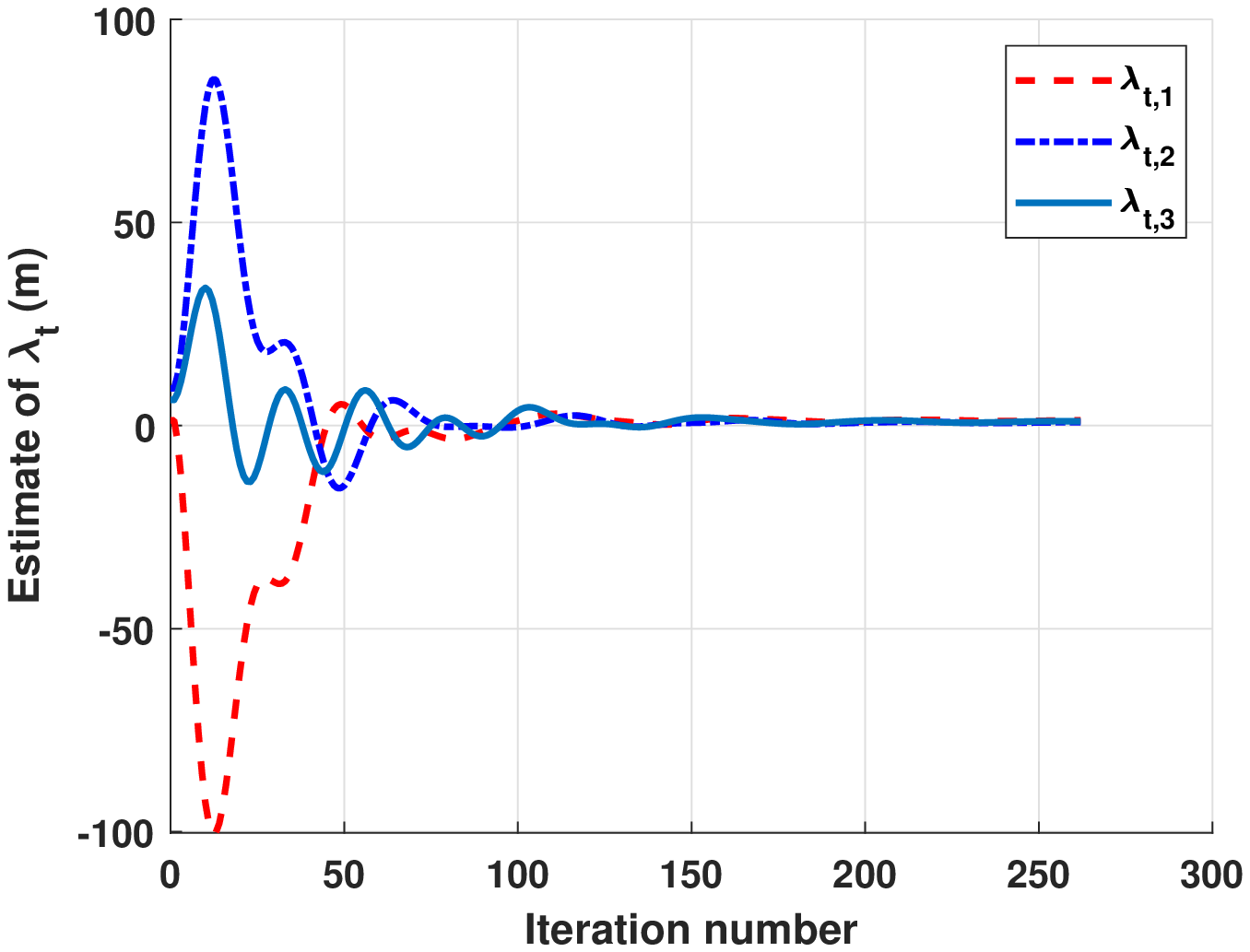}
	}
	\subfigure[]{
		\includegraphics[width=1.625in]{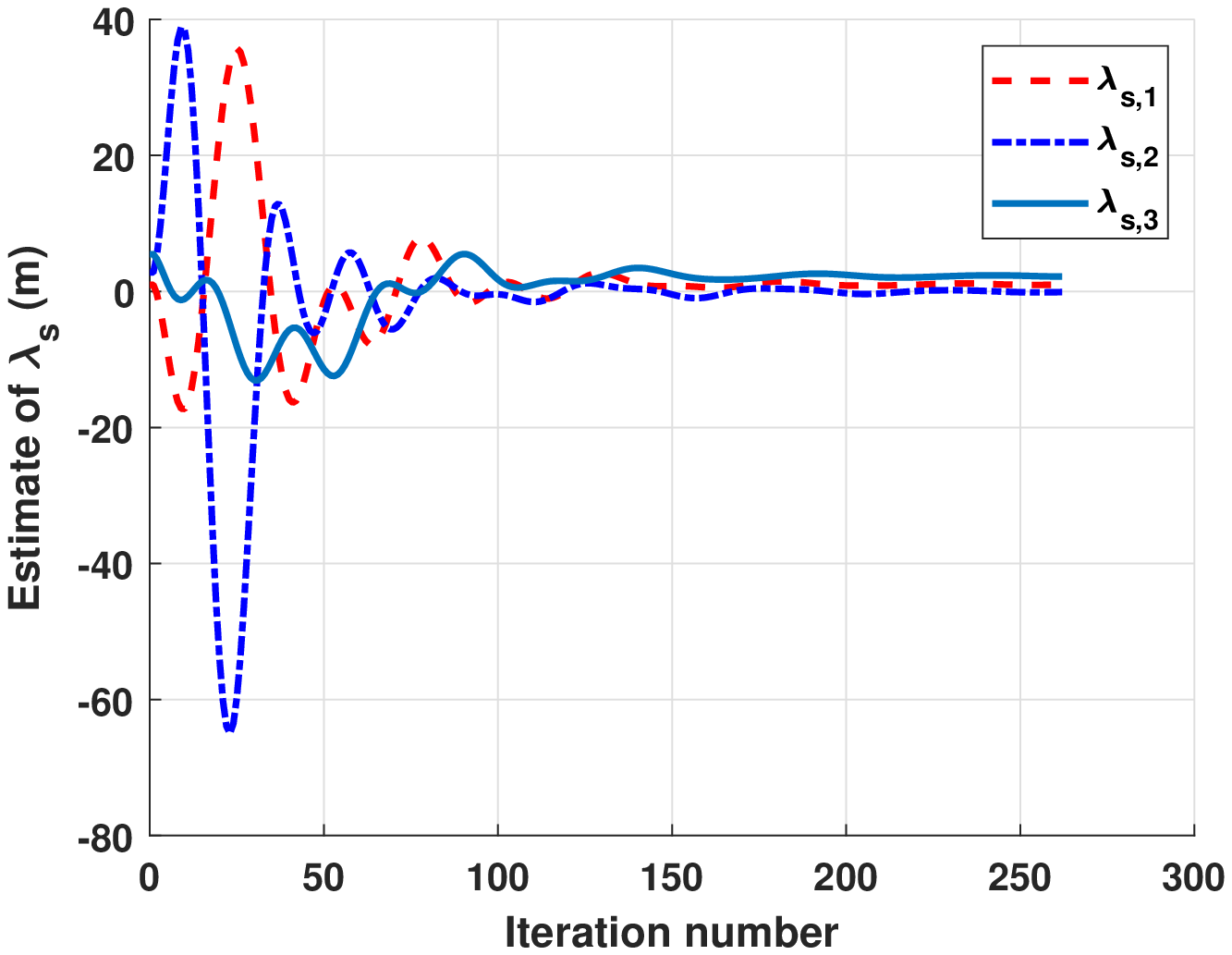}
	}
	\caption{Dynamics and trajectory of estimated parameters of modified LPNN approach when SNR = 20 dB, starting with a randomly initial value. The true target position is $ [50, 50]^\text{T} $. (a) Dynamics of $ \boldsymbol{u} $; (b) Trajectory of $ \boldsymbol{u} $; (c) Dynamics of $ \boldsymbol{\lambda_t} = [\lambda_{t,1}, \lambda_{t,2}, \lambda_{t,3}]^\text{T} $; (d) Dynamics of $ \boldsymbol{\lambda_s} = [\lambda_{s,1}, \lambda_{s,2}, \lambda_{s,3}]^\text{T} $.}
	\label{fig:lpnnw_dynamics}
\end{figure}

\begin{figure}[!t]
	\centering
	\includegraphics[width=3.5in]{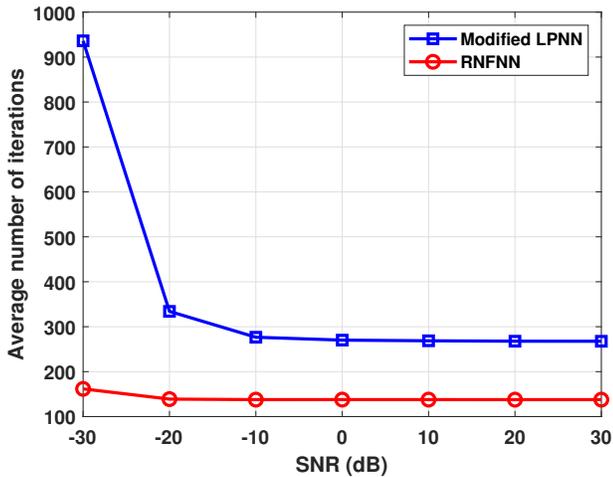}
	\caption{Average number of iterations required to converge for the RNFNN and modified LPNN approaches under different SNRs, where the iterative termination condition $ \epsilon_1 = 10^{-10} $.}
	\label{fig:aveIteNum}
\end{figure}

To further observe the dynamic behavior of the neural networks, two typical examples of the dynamics of RNFNN and modify LPNN are given in Figs. \ref{fig:RNFNN_dynamics} and \ref{fig:lpnnw_dynamics}. The curves show that the RNFNN is more stable and converges faster than the modified LPNN, where the SNR is high. It is worth mentioning that, when the SNR is low, there is a huge difference between the iteration numbers required by the two approaches. For much clear comparison, the average number of iterations required for these two approaches to converge under different SNRs are given in Fig. \ref{fig:aveIteNum}. The setting is described in Section \ref{sec:simulation}.A. It can be seen that for high and moderate SNR levels, both the neural networks converge quickly, and the RNFNN is much faster than the modified LPNN. When the SNR value is very low, the RNFNN preserves its convergence speed, whereas the average number of iterations required by the modified LPNN increases significantly. 

\subsection{Computational Complexity Analysis} \label{subsec_complexity}
\begin{figure}[!t]
	\centering
	\includegraphics[width=3.5in]{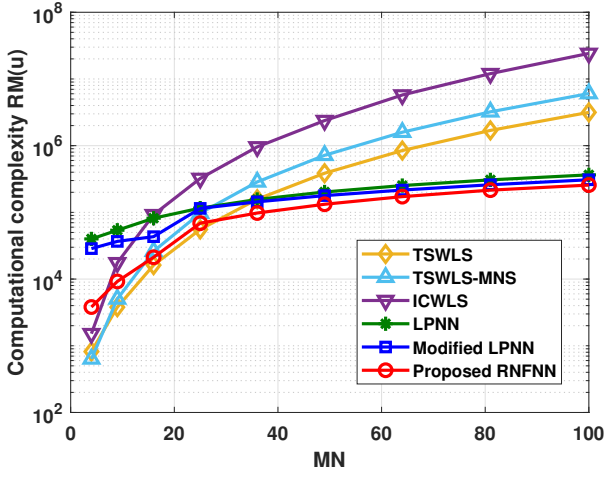}
	\caption{Computational complexity RM($ \boldsymbol{u} $) that illustrates all schemes' complexity shown in Table  \ref{tab:complexity_RMs}, where $ D = 2 $ and SNR = -20dB. }
	\label{fig:difMN_complexity_RM}
\end{figure}

In this subsection, we analyze the computational complexity of the proposed approach and draw a comparison with some state-of-the-art algorithms \cite{2017-Amiri-AsymptoticallyEfficientTarget,2020-Noroozi-ClosedFormSolution,2017-Noroozi-IterativeTargetLocalization,2016-Liang-LagrangeProgrammingNeural}. The number of real multiplications (RMs) involved in computing the parameter $ \boldsymbol{u} $ is represented by RM($ \boldsymbol{u} $) for ease of notation \cite{2020-Noroozi-ClosedFormSolution}. For the analog neural network methods, i.e. the proposed RNFNN, modified LPNN and LPNN methods, the state of all neurons can be updated simultaneously by using parallel processing elements, and the computational complexity per iteration is dominated by their state equations, i.e. (\ref{eq:stateEq_RNFNN}) and (\ref{eq:stateEq_LPNN}). For those methods based on multi-stage weighted least squares (mSWLS) strategies, e.g., the TSWLS, TSWLS-MNS, and ICWLS, due to multiple inversion operations of the covariance matrix involved, their computational complexity, for a large number of transmitters and receivers, increases cubically with factor $ M\!N $ (the product of the number of transmitting and receiving antennas). The total number of RMs required to determine the final estimates of the target location $ \boldsymbol{u} $ with different estimators are tabulated in Table \ref{tab:complexity_RMs}. At the same time, we add a figure that illustrates all schemes' complexity shown in Table \ref{tab:complexity_RMs} for much clear comparison, as shown in Fig. \ref{fig:difMN_complexity_RM}. As reported in table \ref{tab:complexity_RMs}, the modified LPNN estimator requires more $ (M\!N) $ RMs in each iteration than the LPNN in \cite{2016-Liang-LagrangeProgrammingNeural} due to the weighting for different noise variances of transmitter-receiver pairs, however, the latter requires more iterations to converge as shown in Fig. \ref{fig:difMN_complexity_RM}. Although the proposed method is an iterative one, it has a fast convergence speed, as demonstrated in Fig. \ref{fig:aveIteNum}. It is also worth mentioning that the number of iterations required by RNFNN is significantly smaller than that required by (modified) LPNN.


\begin{table}[!t]
	\renewcommand{\arraystretch}{1.0}
	\caption{Comparison of computational complexity RM($ \boldsymbol{u} $)}
	\label{tab:complexity_RMs}
	\centering
	\begin{threeparttable} 
		\begin{tabular}{cc}
			\toprule
			Method & computational complexity RM($ \boldsymbol{u} $) \\
			\midrule
			TSWLS\cite{2017-Amiri-AsymptoticallyEfficientTarget} & \makecell[c]{$ 3(MN)^3+(MN)^2(M\!+\!D)+2MN(M\!+\!D)^2 $ \\ $ +MN(M\!+\!D)+4(M\!+\!D)^3+D(M\!+\!D)^2 $ \\ $ +21(M\!+\!3)+27 $} \\
			\midrule
			TSWLS-MNS\cite{2020-Noroozi-ClosedFormSolution} & \makecell[c]{$ 6(MN)^3+2D(MN)^2+14DMN $ \\ $ +6N+63 $} \\
			\midrule
			ICWLS\cite{2017-Noroozi-IterativeTargetLocalization} & \makecell[c]{$ I_{\text{ICWLS}}\left( (MN)^3+(M\!+\!D)^3+3M^3 \right) $ \\ $ + (MN)^3+(M\!+\!D)^3 $} \\
			\midrule
			LPNN\cite{2016-Liang-LagrangeProgrammingNeural} & $ I_{\text{LPNN}}\left( (3D\!+\!6)(M\!+\!N) \right) $ \\
			\midrule
			Modified LPNN & $ I_{\text{mLPNN}}\left( MN+(3D\!+\!6)(M\!+\!N) \right) $ \\
			\midrule
			Proposed RNFNN & $ I_{\text{RNFNN}}\left( MN+(2D\!+\!4)(M\!+\!N) \right) $ \\
			\bottomrule
		\end{tabular} 
	\end{threeparttable}
	\begin{tablenotes}
		\footnotesize
		\item[] where $ M $ and $ N $ represent the number of transmitters and receivers,  $ D $ denotes localization space dimension ($ D = 2 $ or $ 3 $), and $ I $ with a subscription represents the number of iterations required by the corresponding method. 
	\end{tablenotes} 
\end{table}

\section{Extension to Target Localization with Antenna Position Errors} \label{sec_Extension}
In many practical scenarios, the transmitters and receivers are mounted on moving platforms, and their positions are typically obtained by the use of some methods, such as self-localization \cite{2009-Lui-SemiDefiniteProgramming,2005-Patwari-Locatingnodescooperative} or GPS \cite{2005-Gustafsson-Mobilepositioningusing}. The obtained antenna positions are therefore subject to estimation or measurement errors. To improve the localization performance, these antenna position errors should be taken into account in addition to BR measurement noise. Here, we follow \cite{2014-Rui-EllipticLocalizationPerformance}, the observed transmitter and receiver positions are modeled as 
\begin{equation} \label{eq:APmesurement_model}
\begin{aligned}
\tilde{\boldsymbol{t}}_m & = \boldsymbol{t}_m + \Delta \boldsymbol{t}_m, \quad m = 1,\ldots,M, \\
\tilde{\boldsymbol{s}}_n & = \boldsymbol{s}_n + \Delta \boldsymbol{s}_n, \quad n = 1,\ldots,N.
\end{aligned}
\end{equation}

For notation simplicity, we collect the true antenna positions
into vector $ \boldsymbol{p}= \left[ \boldsymbol{t}^\text{T}, \boldsymbol{s}^\text{T} \right]^\text{T} \in \mathbb{R}^{D(M+N)} $, where $ \boldsymbol{t}= \left[\boldsymbol{t}_1^\text{T},\boldsymbol{t}_2^\text{T},...,\boldsymbol{t}_M^\text{T} \right]^\text{T} $ and  $ \boldsymbol{s}= \left[ \boldsymbol{s}_1^\text{T},\boldsymbol{s}_2^\text{T},...,\boldsymbol{s}_N^\text{T} \right]^\text{T} $. The observed antenna position vector $ \tilde{\boldsymbol{p}} $ and the error vector $ \Delta \boldsymbol{p} $ are defined similarly such that $ \tilde{\boldsymbol{p}}=\boldsymbol{p}+\Delta \boldsymbol{p} $ holds.
$ \Delta \boldsymbol{t}_m $ and $ \Delta \boldsymbol{s}_n $ are assumed independent with each other and modeled as zero-mean Gaussian variables  with variance $ \sigma_{\boldsymbol{t},m}^2 $ and $ \sigma_{\boldsymbol{s},n}^2 $, i.e., 
\begin{equation}
\begin{aligned}
\Delta \boldsymbol{t}_m & \sim \mathcal{N} \left( \boldsymbol{0}_{D \times 1}, \sigma_{\boldsymbol{t},m}^2 \boldsymbol{I}_{D}  \right), \quad m = 1,\ldots,M, \\
\Delta \boldsymbol{s}_n & \sim \mathcal{N} \left( \boldsymbol{0}_{D \times 1}, \sigma_{\boldsymbol{s},n}^2 \boldsymbol{I}_{D}  \right),\quad n = 1,\ldots,N.
\end{aligned}
\end{equation}

We assume that $ \left\lbrace \sigma_{\boldsymbol{t},m}^2 \right\rbrace $ and $ \left\lbrace \sigma_{\boldsymbol{s},n}^2 \right\rbrace $ are known, which is a commonplace assumption in the localization literature (see e.g. \cite{2007HoSourceLocalizationUsing,2019AmiriEfficientEllipticLocalization}). In practice, they can be obtained by measuring the amount of perturbation in the transmitter and receiver positions, during the calibration process via a target with known location.

Combining (\ref{eq:BRmesurement_model}) and (\ref{eq:APmesurement_model}),
the ML estimation problem now is
\begin{equation}  \label{eq:EML} 
\begin{aligned}
\min \limits_{\boldsymbol{u}} \quad \sum_{m=1}^M \sum_{n=1}^N w_{mn} \left(\tilde{r}_{mn}-  \left\| \boldsymbol{u} - \boldsymbol{t}_m \right\|_2 - \left\| \boldsymbol{u} - \boldsymbol{s}_n \right\|_2  \right)^2 \\
+ \sum_{m=1}^M w_{\boldsymbol{t},m} \left\| \tilde{\boldsymbol{t}}_m - \boldsymbol{t}_m \right\|_2^2 + \sum_{n=1}^N w_{\boldsymbol{s},n} \left\| \tilde{\boldsymbol{s}}_n - \boldsymbol{s}_n \right\|_2^2
\end{aligned}
\end{equation}
where the weighting factor
\begin{equation}
\begin{aligned}
w_{\boldsymbol{t},m} = \dfrac{1/\sigma_{\boldsymbol{t},m}^2}{\sum_{m=1}^M \{1/\sigma_{\boldsymbol{t},m}^2 \}}
\end{aligned}
\end{equation}
and
\begin{equation}  
w_{\boldsymbol{s},n} = \dfrac{1/\sigma_{\boldsymbol{s},n}^2}{\sum_{n=1}^N \{1/\sigma_{\boldsymbol{s},n}^2 \}}.
\end{equation}
Note that, when there are no information of $ \left\lbrace \sigma_{\boldsymbol{t},m}^2 \right\rbrace $ and $ \left\lbrace \sigma_{\boldsymbol{s},n}^2 \right\rbrace $, we can simply set $ w_{\boldsymbol{t},m}=w_{\boldsymbol{s},n}=1 $ for all $ m $ and $ n $.

Similarly, to solve the nonlinear optimization problem (\ref{eq:EML}), an extended RNF can be constructed as
\begin{equation} \label{eq:ERNF}
\begin{aligned}
E(\boldsymbol{x}) = & \dfrac{1}{2} \sum_{m=1}^M \sum_{n=1}^N w_{mn} \left( \tilde{r}_{mn} - h_{\boldsymbol{t},m} - h_{\boldsymbol{s},n} \right) ^2 \\
& + \sum_{m=1}^M w_{\boldsymbol{t},m} \left\| \tilde{\boldsymbol{t}}_m - \boldsymbol{t}_m \right\|_2^2 + \sum_{n=1}^N w_{\boldsymbol{s},n} \left\| \tilde{\boldsymbol{s}}_n - \boldsymbol{s}_n \right\|_2^2 \\
& + \dfrac{\rho}{4} \sum_{m=1}^M \left( h_{\boldsymbol{t},m}^2 -\left\| \boldsymbol{u} - \boldsymbol{t}_m \right\|_2^2 \right)  ^2 \\
& + \dfrac{\rho}{4} \sum_{n=1}^N \left(  h_{\boldsymbol{s},n}^2 -\left\| \boldsymbol{u} - \boldsymbol{s}_n \right\|_2^2 \right)  ^2 . \\
\end{aligned}
\end{equation}
Note that the length of $ \boldsymbol{x} = \left[  \boldsymbol{u}^{\text{T}},  \boldsymbol{p}^{\text{T}}, \boldsymbol{h}^{\text{T}}_{\boldsymbol{t}}, \boldsymbol{h}^{\text{T}}_{\boldsymbol{s}} \right] ^{\text{T}} $ is now $ \left( D+\left( D+1\right) \left( M+N\right) \right) $, because of additional unknown antenna position vector $ \boldsymbol{p} $. The dynamics of $ \boldsymbol{u} $, $ \left\lbrace h_{\boldsymbol{t},m} \right\rbrace  $ and $ \left\lbrace h_{\boldsymbol{s},n} \right\rbrace  $ are the same as in (\ref{eq:stateEq_RNFNN}). For the
other states, their dynamics are described as follows. 
\begin{equation} \label{eq:stateEq_p}
\left\{
\begin{aligned}
\dfrac{d \boldsymbol{t}_m }{dt} = &  w_{\boldsymbol{t},m} \left( \tilde{\boldsymbol{t}}_m - \boldsymbol{t}_m \right) \\ 
& -\rho \left( h_{\boldsymbol{t},m}^2 -\left\| \boldsymbol{u} - \boldsymbol{t}_m \right\|_2^2 \right) \left( \boldsymbol{u} - \boldsymbol{t}_m \right) , \\
&  \quad m = 1,\ldots,M, \\
\dfrac{d \boldsymbol{s}_n }{dt} = &  w_{\boldsymbol{s},n} \left(  \tilde{\boldsymbol{s}}_n - \boldsymbol{s}_n \right) \\
& -\rho \left( h_{\boldsymbol{s},n}^2 -\left\| \boldsymbol{u} - \boldsymbol{s}_n \right\|_2^2 \right) \left( \boldsymbol{u} - \boldsymbol{s}_n \right) , \\
&  \quad n = 1,\ldots,N.
\end{aligned}	
\right.
\end{equation}

\begin{algorithm}[!t]
	\caption{Relaxed Energy Function based Neural Network Approach to Localization Without Antenna Position Errors}
	\label{alg:RNFNN}
	\begin{algorithmic}[]
		\Require
		Transmitter and receiver positions $ \{ \boldsymbol{t}_m \} $, $ \{ \boldsymbol{s}_n \} $ and BR measurements $ \left\lbrace \tilde{r}_{mn}\right\rbrace $.		
		\Ensure
		Target location estimate $ \hat{\boldsymbol{u}} $.
		\State [Step 1] Initialization:
		\begin{itemize}
			\item Let $ t=0 $ and take parameter $ \rho > 0 $.
			\item Choose arbitrarily initial point $ \boldsymbol{x}(0)= \boldsymbol{x}^0 $. 
			\item Take the step size $ \Delta t > 0 $, iterative termination condition $ \epsilon_1 $, maximum number of iterations $ T $ and $ k=1 $. 
		\end{itemize}
		\State [Step 2] Computation of gradient:
		\begin{itemize}
			\item Compute the transient behaviors by (\ref{eq:stateEq_RNFNN}).  
			\item Set $ \boldsymbol{a}(t)= \dot{\boldsymbol{x}} = \left[ \frac{d\boldsymbol{u}}{dt}, \frac{d\boldsymbol{h_t}}{dt}, \frac{d\boldsymbol{h_s}}{dt} \right]^\text{T} $.
		\end{itemize}
		\State [Step 3] States updating:
		\begin{equation*}
		\left\{
		\begin{aligned}
		\boldsymbol{u}(t+\Delta t) & = \boldsymbol{u}(t) + \Delta t  \frac{d\boldsymbol{u}}{dt} \\
		\boldsymbol{h_t}(t+\Delta t) & = \boldsymbol{h_t}(t) + \Delta t \frac{d\boldsymbol{h_t}}{dt} \\
		\boldsymbol{h_s}(t+\Delta t) & = \boldsymbol{h_s}(t) + \Delta t \frac{d\boldsymbol{h_s}}{dt}
		\end{aligned}	
		\right.
		\end{equation*}
		\State [Step 4] Calculation:
		\begin{itemize}
			\item $ e = \left\| \boldsymbol{a}(t) \right\|_2^2 = \sum_{i=1}^{D+M+N} \boldsymbol{a}_i^2(t) $.
		\end{itemize}
		\State [Step 5] Feedback rule:
		\begin{itemize}
			\item If $ e < \epsilon_1 $ or $ k \geq T $, then stop, output $ \hat{\boldsymbol{u}} = \boldsymbol{u}(t) $.
			\item Otherwise, set $ k=k+1 $ and go to step 2.
		\end{itemize}
	\end{algorithmic}
\end{algorithm}

\begin{algorithm}[!t]
	\caption{Relaxed Energy Function based Neural Network Approach to Localization With Antenna Position Errors}
	\label{alg:RNFNN_P}
	\begin{algorithmic}[]
		\Require
		Observed transmitter and receiver positions $ \{ \tilde{\boldsymbol{t}}_m \} $, $ \{ \tilde{\boldsymbol{s}}_n \} $ and BR measurements $ \left\lbrace \tilde{r}_{mn}\right\rbrace $.		
		\Ensure
		Target location estimate $ \hat{\boldsymbol{u}} $.
		\State [Step 1] Initialization:
		\begin{itemize}
			\item Let $ t=0 $ and take parameter $ \rho > 0 $.
			\item Choose arbitrarily initial point $ \boldsymbol{x}(0)= \boldsymbol{x}^0 $. 
			\item Take the step size $ \Delta t > 0 $, iterative termination condition $ \epsilon_1 $, maximum number of iterations $ T $ and $ k=1 $. 
		\end{itemize}
		\State [Step 2] Computation of gradient:
		\begin{itemize}
			\item Compute the transient behaviors by (\ref{eq:stateEq_RNFNN}) and (\ref{eq:stateEq_p}).  
			\item Set $ \boldsymbol{a}(t)= \dot{\boldsymbol{x}} = \left[ \frac{d\boldsymbol{u}}{dt}, \frac{d\boldsymbol{t}}{dt}, \frac{d\boldsymbol{s}}{dt}, \frac{d\boldsymbol{h_t}}{dt}, \frac{d\boldsymbol{h_s}}{dt} \right]^\text{T} $.
		\end{itemize}
		\State [Step 3] States updating:
		\begin{equation*}
		\left\{
		\begin{aligned}
		\boldsymbol{u}(t+\Delta t) & = \boldsymbol{u}(t) + \Delta t  \frac{d\boldsymbol{u}}{dt} \\
		\boldsymbol{t}(t+\Delta t) & = \boldsymbol{t}(t) + \Delta t  \frac{d\boldsymbol{t}}{dt} \\
		\boldsymbol{s}(t+\Delta t) & = \boldsymbol{s}(t) + \Delta t  \frac{d\boldsymbol{s}}{dt} \\
		\boldsymbol{h_t}(t+\Delta t) & = \boldsymbol{h_t}(t) + \Delta t \frac{d\boldsymbol{h_t}}{dt} \\
		\boldsymbol{h_s}(t+\Delta t) & = \boldsymbol{h_s}(t) + \Delta t \frac{d\boldsymbol{h_s}}{dt}
		\end{aligned}	
		\right.
		\end{equation*}
		\State [Step 4] Calculation:
		\begin{itemize}
			\item $ e = \left\| \boldsymbol{a}(t) \right\|_2^2 = \sum_{i=1}^{D+M+N} \boldsymbol{a}_i^2(t) $.
		\end{itemize}
		\State [Step 5] Feedback rule:
		\begin{itemize}
			\item If $ e < \epsilon_1 $ or $ k \geq T $, then stop, output $ \hat{\boldsymbol{u}} = \boldsymbol{u}(t) $.
			\item Otherwise, set $ k=k+1 $ and go to step 2.
		\end{itemize}
	\end{algorithmic}
\end{algorithm}

\section{Simulation Results and Discussions} \label{sec:simulation}
In this section, simulations are conducted in several different scenarios to evaluate the performance of the proposed method. In the first two scenarios, the localization without antenna position errors is considered in 2-dimension (2-D) and 3-dimension (3-D) space, respectively. Then in the third scenario, we evaluate the proposed method for different possible target locations. Next, we investigate the positioning accuracy for radars with the different numbers of antennas and analyze the computational complexity of the proposed method via CPU running time.
At last two scenarios, the antenna position errors are taken into account in addition to BR measurement noise. 
The performance of the proposed estimator is compared with some state-of-the-art methods, including the two-stage weighted least squares (TSWLS) method in \cite{2017-Amiri-AsymptoticallyEfficientTarget}, the two-stage weighted least squares with minimum number of sensors (TSWLS-MNS) method in \cite{2020-Noroozi-ClosedFormSolution}, the iterative constrained weighted least squares (ICWLS) method in \cite{2017-Noroozi-IterativeTargetLocalization} and the Lagrange programming neural network (LPNN) method in \cite{2016-Liang-LagrangeProgrammingNeural}. They were shown to outperform other existing methods. Furthermore, the root of CRLB (root CRLB) is consider as a benchmark, which provides a lower bound on the accuracy of any unbiased estimators \cite{1993-Kay-FundamentalsStatisticalSignal}. 

Here we follow \cite{2003-Leung-highperformancefeedback}, the analog computation of Fig. \ref{fig:diagram_RNFNN} is realized in a discrete manner by the Euler method for the convenience to observe the dynamic behavior of neural networks, instead of using MATLAB ODE solver. They are summarized in Algorithm 1 and Algorithm 2, respectively. The LPNN counterpart is implemented in a similar manner. To start the procedure, the initial value $ \boldsymbol{u}^0=[x^0, y^0, z^0]^\text{T} m $ is randomly selected according to uniform distribution as $ x^0, y^0, z^0 \sim \mathcal{U} \left( -400,400 \right) m $, and the value of $ \left\lbrace h_{\boldsymbol{t},m}^0 \right\rbrace $ and $ \left\lbrace h_{\boldsymbol{s},n}^0 \right\rbrace $ are determined according to (\ref{eq:gt}) and (\ref{eq:gs}). The $ \left\lbrace \lambda_{\boldsymbol{t},m} \right\rbrace $ and $ \left\lbrace \lambda_{\boldsymbol{s},n} \right\rbrace $ are initialized as uniform
numbers distributed between 0 and 1. In this study, $ \rho=0.1 $ and $ C=1 $ are employed. In addition, we set the iterative termination condition $ \epsilon_1 = 10^{-10} $ and maximum number of iterations $ T $ sufficiently large.

From the practical point of view, the noise in BR measurements depends on
the SNR of each transmitter-receiver pair \cite{1984-Barton-HandbookRadarMeasurement}. Here, similar to \cite{2017-Amiri-AsymptoticallyEfficientTarget,2020-Noroozi-ClosedFormSolution}, the variance of the BR noise corresponding to the $ m $-th transmitter and $ n $-th receiver, is considered as $ \sigma_{mn}^2 = k_1 g_{\boldsymbol{t},m}^2 g_{\boldsymbol{s},n}^2 $, and define the average SNR \cite{2005-Willis-BistaticRadar,2016-He-GeneralizedCramerRao}
\begin{equation}
\begin{aligned}
SNR & = 10 \text{log}_{10} \left( \frac{1}{MN} \sum_{m=1}^M \sum_{n=1}^N \frac{k_2}{\sigma_{mn}^2} \right) \\
& = 10 \text{log}_{10} \left( \frac{1}{MN} \sum_{m=1}^M \sum_{n=1}^N \frac{K}{g_{\boldsymbol{t},m}^2 g_{\boldsymbol{s},n}^2} \right)
\end{aligned}
\end{equation}
where $ K= {k_2}/{k_1} $, $ k_1 $ depends on systematic parameters and target cross section \cite{2005-Willis-BistaticRadar}, and $ k_2 $ is a scale factor. Here, we set $ k_2 = 1000 $ and scale $ k_1 $ to produce different SNRs. The localization performance is evaluated in terms of RMSE and bias of the target location estimate, as well as empirical CDF of positioning error, which are respectively defined as \cite{2013SoSimpleFormulaeBias,2019WangConvexRelaxationMethods}
\begin{equation}
\text{RMSE}(\hat{\boldsymbol{u}}) = \sqrt{ \frac{1}{L} \sum_{l=1}^{L} \left\|  \hat{\boldsymbol{u}}^{(l)} - \boldsymbol{u} \right\|_2^2 } 
\end{equation}
\begin{equation}
\text{Bias}(\hat{\boldsymbol{u}}) = \left\|  \frac{1}{L}\sum_{l=1}^{L}  \hat{\boldsymbol{u}}^{(l)} - \boldsymbol{u} \right\|_1
\end{equation}
\begin{equation}
\text{CDF}(\Delta \boldsymbol{u}) = P \left( \left\|  \hat{\boldsymbol{u}}^{(l)} - \boldsymbol{u} \right\|_2 \leq \Delta \boldsymbol{u} \right) 
\end{equation}
where $ \hat{\boldsymbol{u}}^{(l)} $ denotes the estimate of $ \boldsymbol{u} $ for the $ l $-th trial and $ L $ is the number of Monte Carlo trials which is 5000 in all simulation scenarios.

\begin{table}[!t]
	\caption{Transmitter and Receiver Positions in the 2D Scenario (in Meters) }
	\label{tab:2D_position}
	\centering 
	\begin{tabular}{cccccc}
		\toprule
		Tx no.$ m $ & $ x_{\boldsymbol{t}_m} $ & $ y_{\boldsymbol{t}_m} $ & Rx no.$ n $ & $ x_{\boldsymbol{s}_n} $ & $ y_{\boldsymbol{s}_n} $ \\
		\midrule
		$ 1 $ & $ -1000 $ & $ -1300 $ & $ 1 $ & $ 1500 $ & $ -1800 $ \\
		\midrule
		$ 2 $ & $ 500 $ & $ 2000 $ & $ 2 $ & $ 2100 $ & $ 1500 $ \\
		\midrule
		$ 3 $ & $ 2500 $ & $ 0 $ & $ 3 $ & $ -1200 $ & $ 1000 $ \\
		\bottomrule
	\end{tabular}
\end{table}

\begin{table*}[!t]
	\caption{Transmitter and Receiver Positions in the 3D Scenario (in Meters) }
	\label{tab:3D_position}
	\centering 
	\begin{tabular}{cccccccc}
		\toprule
		Tx no.$ m $ & $ x_{\boldsymbol{t}_m} $ & $ y_{\boldsymbol{t}_m} $ & $ z_{\boldsymbol{t}_m} $ & Rx no.$ n $ & $ x_{\boldsymbol{s}_n} $ & $ y_{\boldsymbol{s}_n} $ & $ z_{\boldsymbol{s}_n} $ \\
		\midrule
		$ 1 $ & $ 2000 $ & $ 3000 $ & $ 800 $ & $ 1 $ & $ 4000 $ & $ 4000 $ & $ 1000 $ \\
		\midrule
		$ 2 $ & $ 2000 $ & $ -3000 $ & $ 1200 $ & $ 2 $ & $ -4500 $ & $ 5000 $ & $ 1500 $ \\
		\midrule
		$ 3 $ & $ -2000 $ & $ 3000 $ & $ 1000 $ & $ 3 $ & $ -4500 $ & $ -4500 $ & $ 1000 $ \\
		\midrule
		$ 4 $ & $ -2000 $ & $ -3000 $ & $ 1600 $ & $ 4 $ & $ 0 $ & $ 6000 $ & $ 1200 $ \\
		\midrule
		$ 5 $ & $ 0 $ & $ 0 $ & $ 1500 $ & $ 5 $ & $ 0 $ & $ -6000 $ & $ 1000 $ \\
		\midrule
		$ - $ & $ - $ & $ - $ & $ - $ & $ 6 $ & $ -6000 $ & $ 0 $ & $ 1000 $ \\
		\bottomrule
	\end{tabular}
\end{table*}

\subsection{Target Localization Without Antenna Position Errors} \label{subsec:A}
In the first two scenarios, we evaluate the performance of the approaches versus SNR, for two cases: 2-dimension (2-D) and 3-dimension (3-D) localization, respectively.

\begin{figure}[!t]
	\centering
	\includegraphics[width=3.5in]{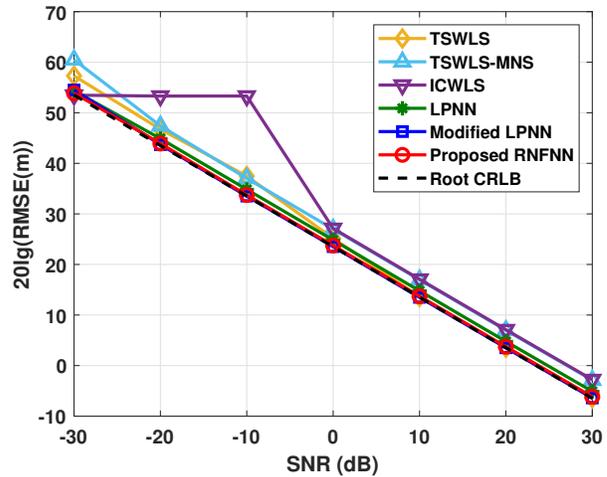}
	\caption{RMSE($ \hat{\boldsymbol{u}} $) versus SNR, where the proposed method is compared with the state-of-the-art algorithms and the root CRLB in the 2-D scenario.}
	\label{fig:2D_rmse}
\end{figure}

\begin{figure}[!t]
	\centering
	\includegraphics[width=3.5in]{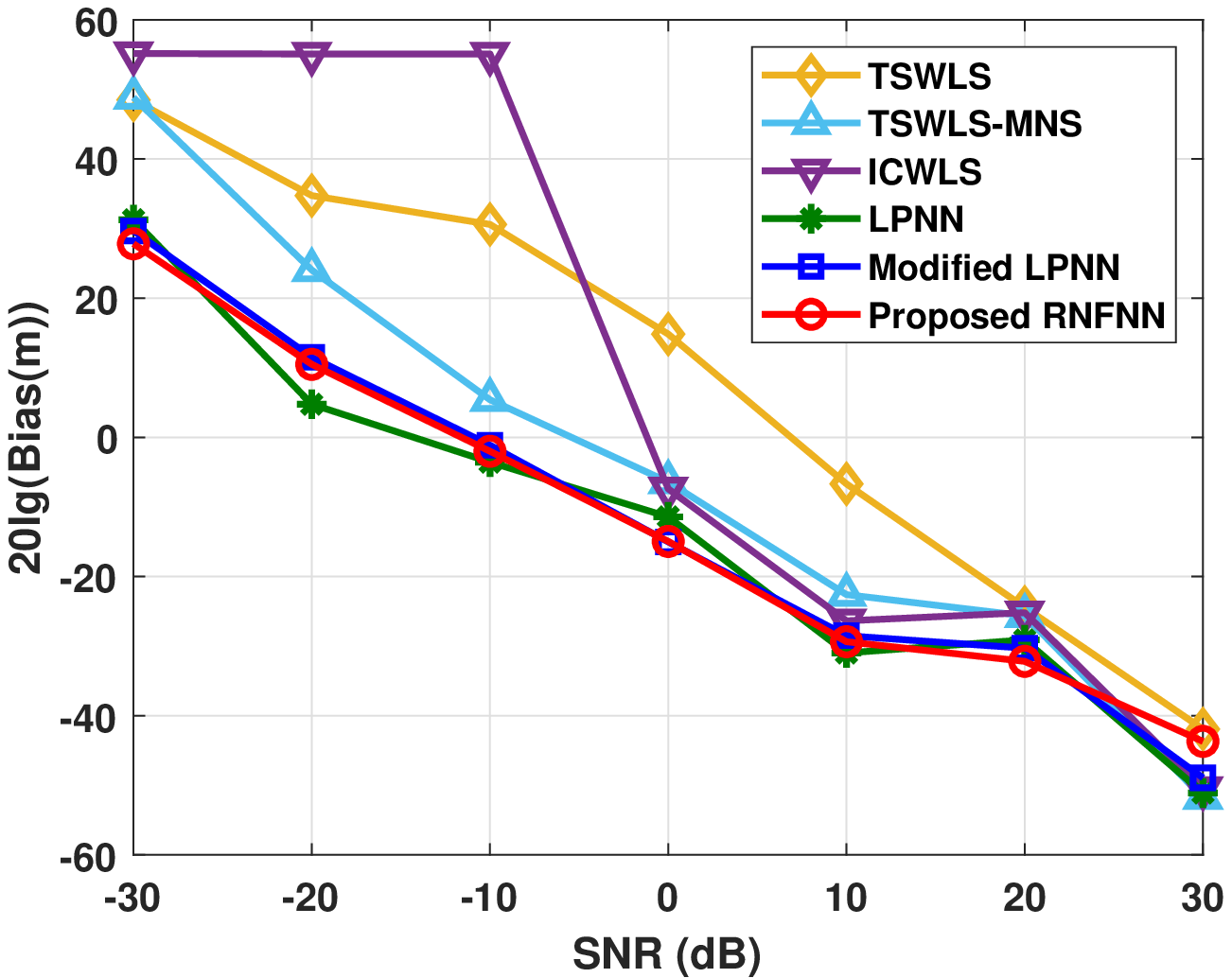}
	\caption{Bias($ \hat{\boldsymbol{u}} $) versus SNR, where the proposed method is compared with the state-of-the-art algorithms in the 2-D scenario.}
	\label{fig:2D_Bias}
\end{figure}

\begin{figure}[!t]
	\centering
	\includegraphics[width=3.5in]{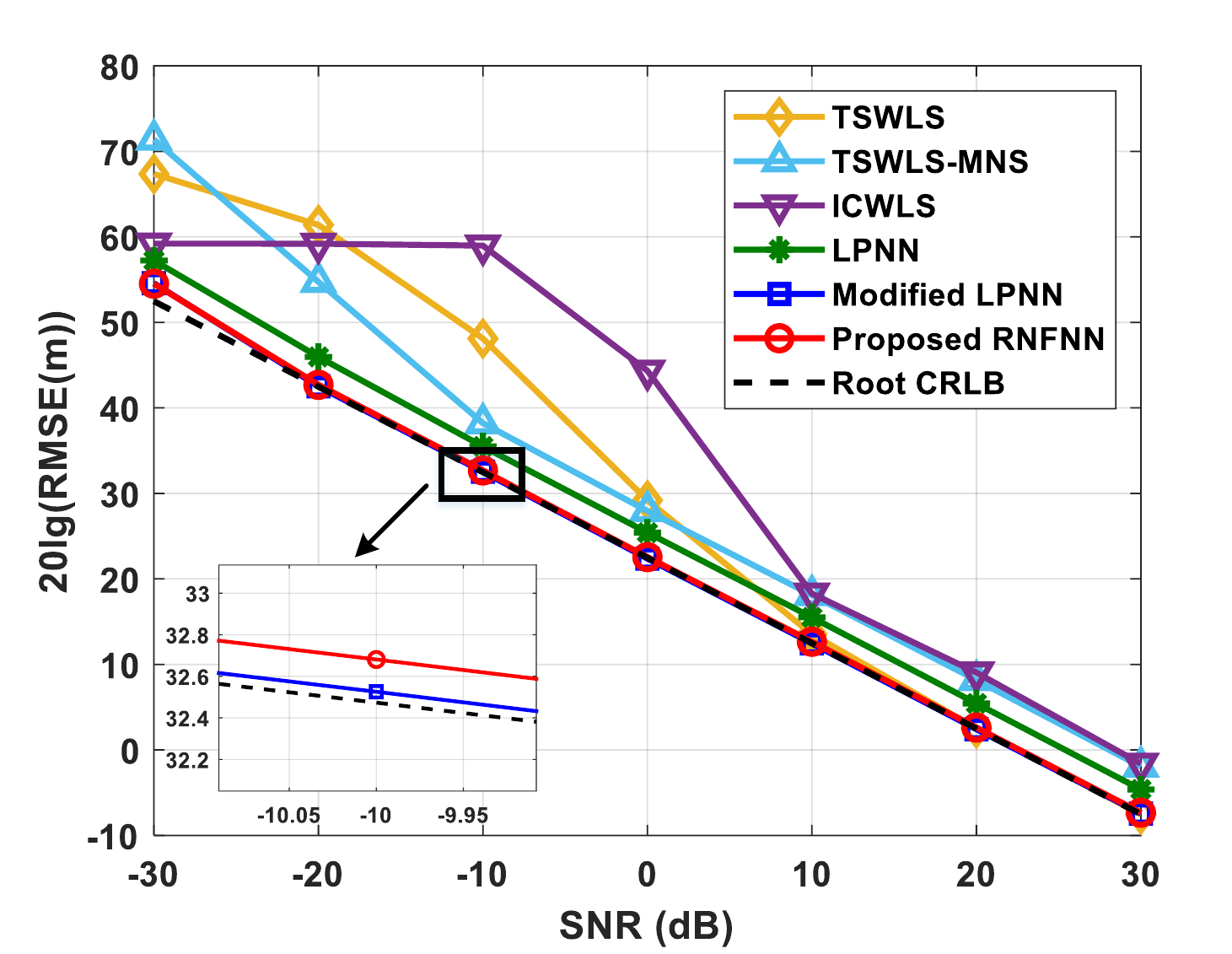}
	\caption{RMSE($ \hat{\boldsymbol{u}} $) versus SNR, where the proposed method is compared with the state-of-the-art algorithms and the root CRLB in the 3-D scenario.}
	\label{fig:3D_rmse}
\end{figure}

\textit{Scenario 1:} We first consider the localization using a distributed MIMO radar with $ M=3 $ transmitters and $ N=3 $ receivers in a 2-D plane, where their positions are accurately known and tabulated in Table \ref{tab:2D_position}. The target is assumed to be located at $ \boldsymbol{u}=[50,50]^\text{T} m $. The SNR varies from $ -30 $ dB to $ 30 $ dB to cover the low, moderate and high SNR conditions. In Figs. \ref{fig:2D_rmse} and \ref{fig:2D_Bias}, the RMSE and bias of the proposed method are compared with the state-of-the-art algorithms. As shown in Fig. \ref{fig:2D_rmse}, both the proposed RNFNN and modified LPNN methods perform well such that they can reach the CRLB accuracy over the high, moderate and low SNR levels. Moreover, they are shown to outperform the other existing methods under the more general noise cases where different transmit-receive pairs have different noise variances. Note that the other existing methods are shown to be able to attain the CRLB at high SNR conditions, under the assumption of equal BR measurement noise variances (when SNR $ \geq $ 0 dB, the TSWLS method can also achieve the CRLB under this general noise model, as shown in Fig. \ref{fig:2D_rmse} and \ref{fig:3D_rmse}). From Fig. \ref{fig:2D_Bias}, we can see that the proposed method has smaller bias, compared to the state-of-the-art methods, in a wide range of SNR levels. 

\begin{figure}[!t]
	\centering
	\includegraphics[width=3.5in]{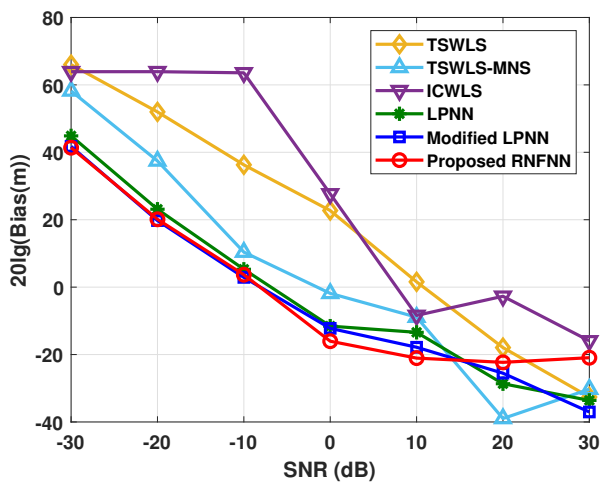}
	\caption{Bias($ \hat{\boldsymbol{u}} $) versus SNR, where the proposed method is compared with the state-of-the-art algorithms in the 3-D scenario.}
	\label{fig:3D_Bias}
\end{figure}

\textit{Scenario 2:} In this scenario, we consider a distributed MIMO radar consisting of $ M = 5 $ transmitters and $ N = 6 $ receivers, whose true positions in a 3-D space are tabulated in Table \ref{tab:3D_position}. The location of target is assumed at $ \boldsymbol{u}=[-500, 600, 550]^\text{T} m $. The corresponding positioning RMSE and bias are given in Figs. \ref{fig:3D_rmse} and \ref{fig:3D_Bias}, respectively. The results are similar to the 2-D case, which show the superior performance of the proposed estimator, in terms of RMSE and bias, compared to the state-of-the-art methods. The proposed one and modified LPNN method has similar performance. In addition, we see here that the performance of the LPNN method is more pronounced than its performance degradation in the 2-D localization scenario.
Furthermore, the well-known threshold phenomenon that usually occurs in methods based on pseudo-linear equations, e.g. the TSWLS, TSWLS-MNS and ICWLS, is observed, where the positioning performance suddenly deviates from CRLB when the SNR levels are lower. Specifically, the RMSE's of the TSWLS and ICWLS significantly deviate from the root CRLB at SNR = 10 dB, and that of the TSWLS-MNS appears at SNR = -10 dB.

\textit{Scenario 3:} In this scenario, we aim to assess the performance of the proposed method for different possible locations of the target in the area-of-interest. The target location, $ \boldsymbol{u}=[x, y, z]^\text{T} m $, in each Monte Carlo trial is randomly selected according to uniform distribution as $ x \sim \mathcal{U} \left( -1000,1000 \right) m $, $ y \sim \mathcal{U} \left( -1000,1000 \right) m $ and $ z \sim \mathcal{U} \left( 0,1000 \right) m $. The antenna positions in a 3-D space are tabulated in Table \ref{tab:3D_position}. 

\begin{figure*}[!t]
	\centering
	\subfigure[]{
		\includegraphics[width=2.25in]{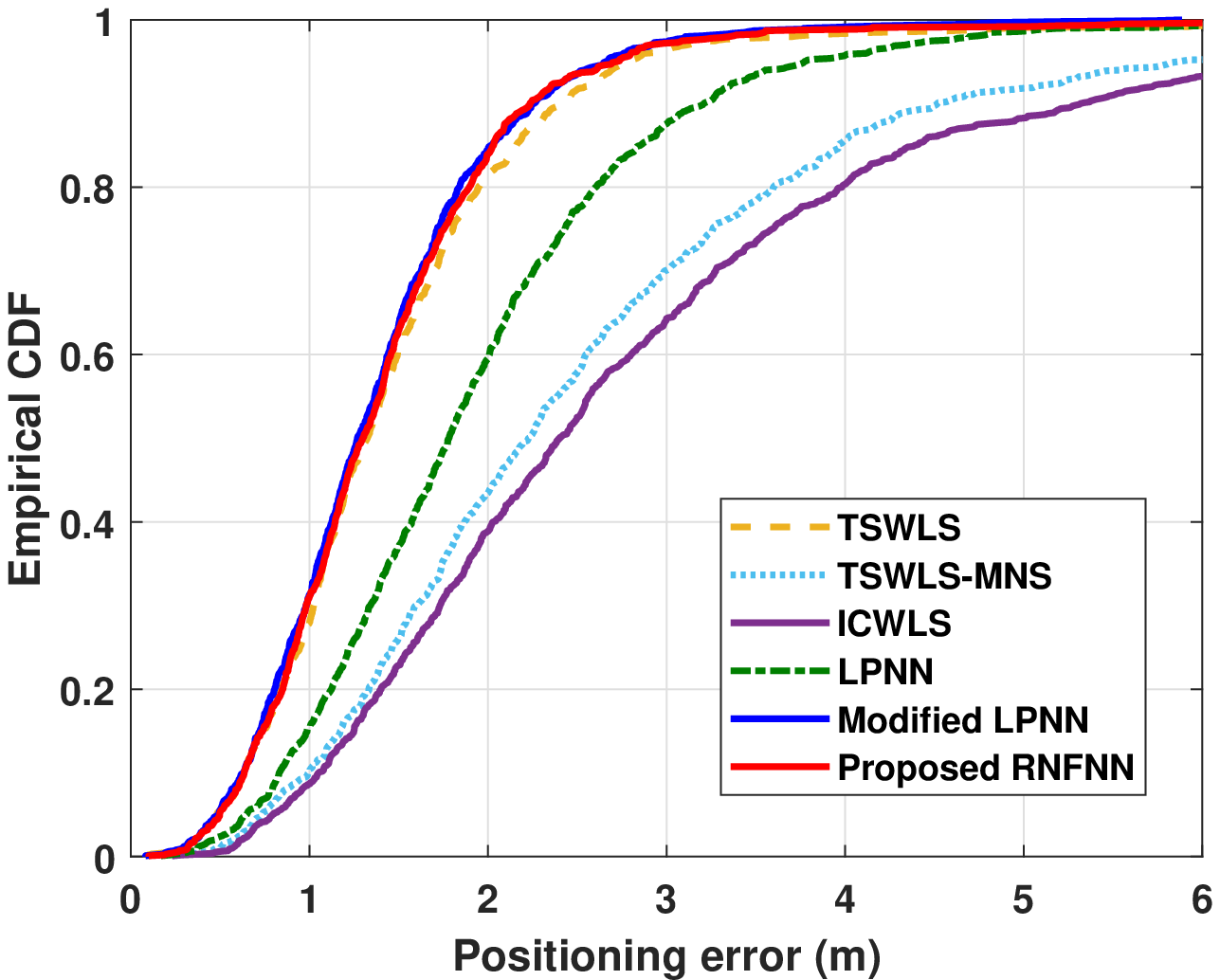}
	}
	\subfigure[]{
		\includegraphics[width=2.25in]{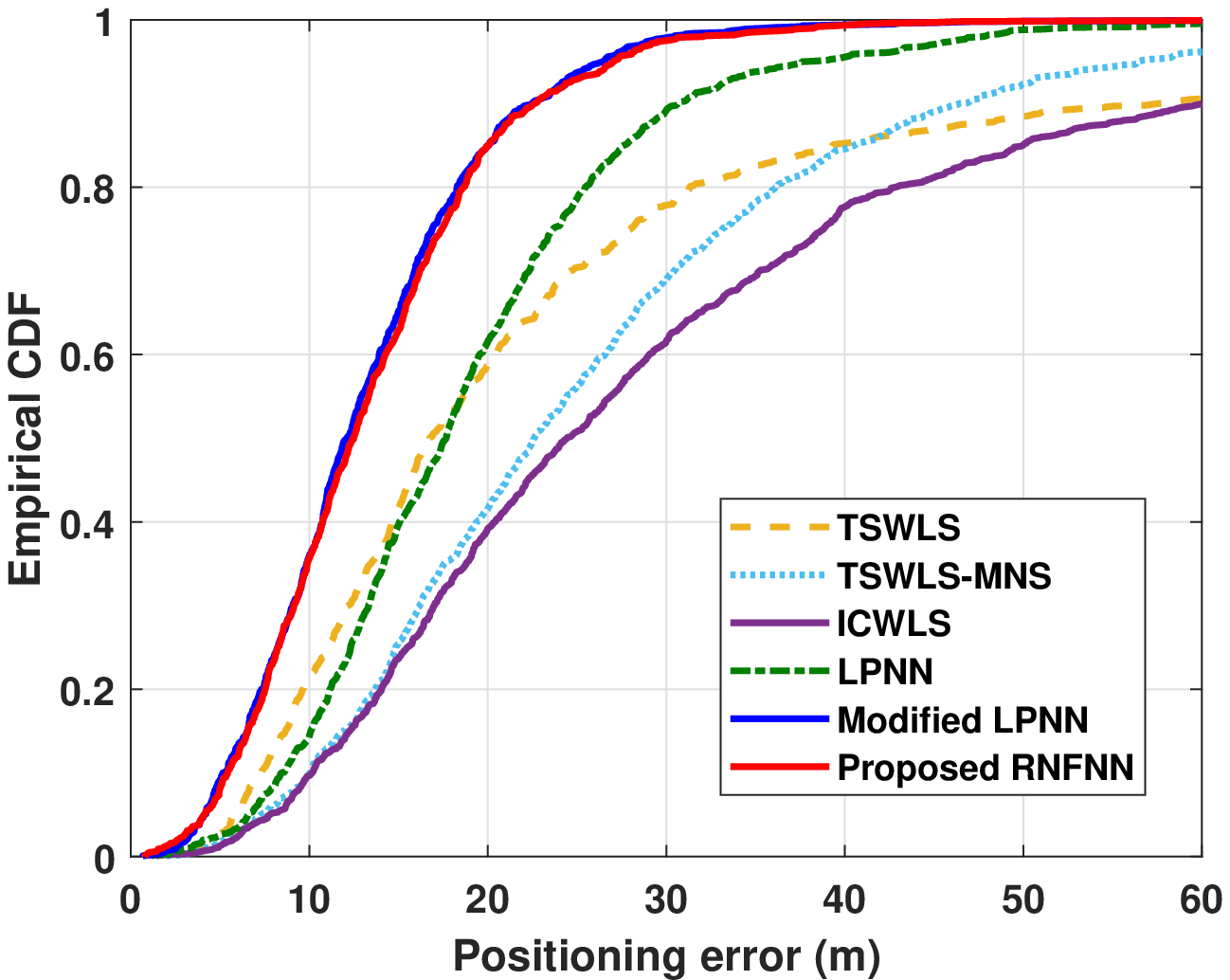}
	}
	\subfigure[]{
		\includegraphics[width=2.25in]{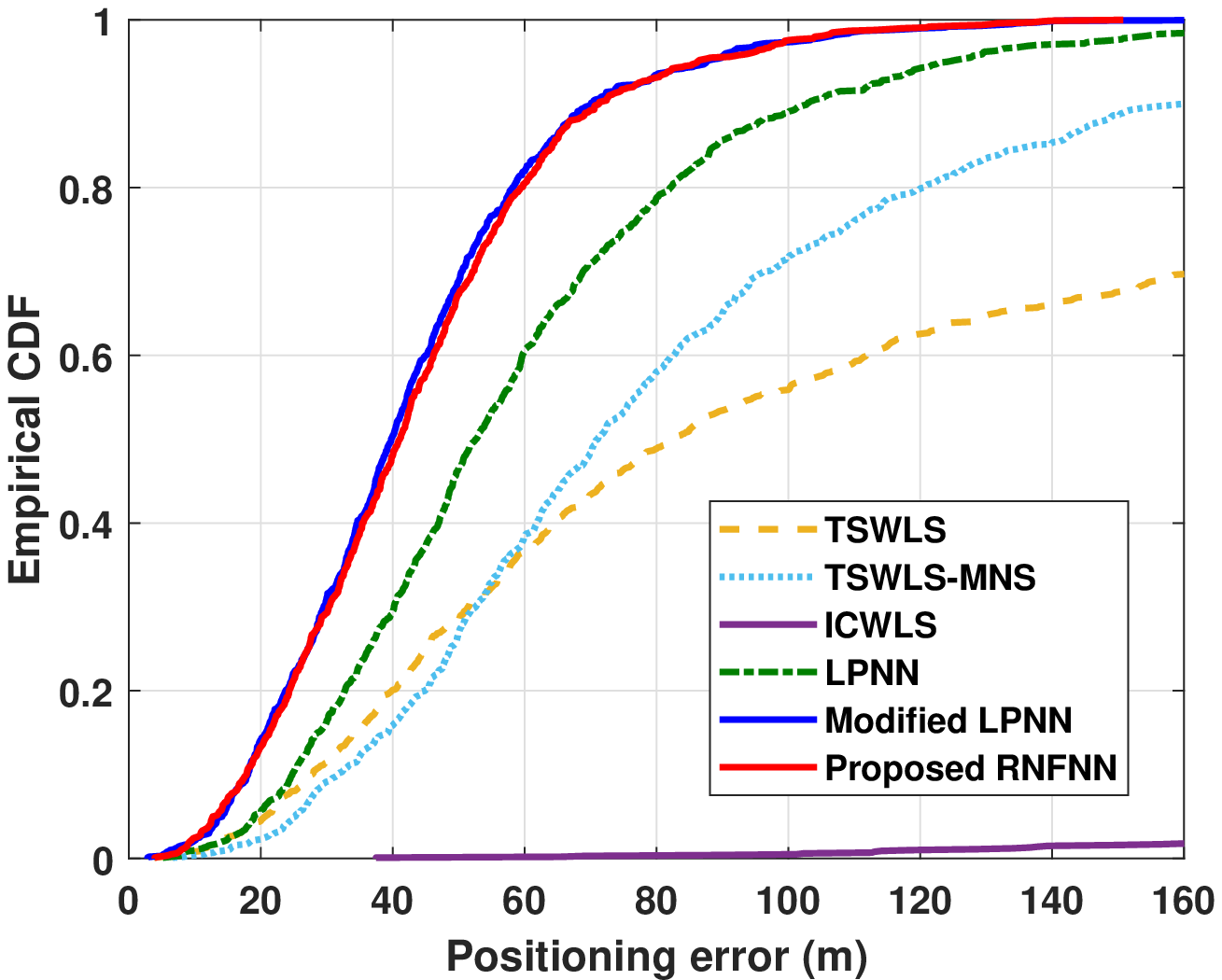}
	}
	\caption{Empirical CDF of positioning error, where the proposed method is compared with the state-of-the-art algorithms (a) high SNR = 20 dB, (b) moderate  SNR = 0 dB, and (c) relatively low SNR = -10 dB.}
	\label{fig:3D_ECDF}
\end{figure*}

In Fig. \ref{fig:3D_ECDF}, we analyze the accuracy of the localization methods by empirical CDF of positioning error, which is evaluated for different SNR levels: high SNR = 20 dB, moderate SNR = 0 dB, and relatively low SNR = -10 dB. As shown in Fig. \ref{fig:3D_ECDF}(a), when the SNR is high, the proposed RNFNN, modified LPNN and TSWLS have similar accuracy, while the TSWLS-MNS, ICWLS and LPNN have more degraded performance. The proposed RNFNN and modified LPNN methods preserve their optimum performance in the moderate SNR level, whereas an obvious degradation occurs in the TSWLS, as depicted in Fig. \ref{fig:3D_ECDF}(b). It can be seen in Fig. \ref{fig:3D_ECDF}(c) that the TSWLS and especially the ICWLS occur severe performance degradation in the relatively low SNR level.

\begin{figure}[!t]
	\centering
	\includegraphics[width=3.5in]{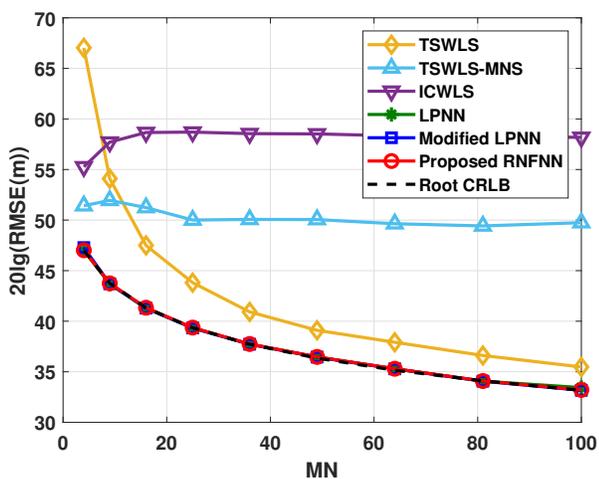}
	\caption{ RMSE($ \hat{\boldsymbol{u}} $) versus $ M\!N $, where the proposed method is compared with the state-of-the-art algorithms and averaged CRLB for SNR = -20 dB. }
	\label{fig:difMN_rmse}
\end{figure}

\begin{figure}[!t]
	\centering
	\includegraphics[width=3.5in]{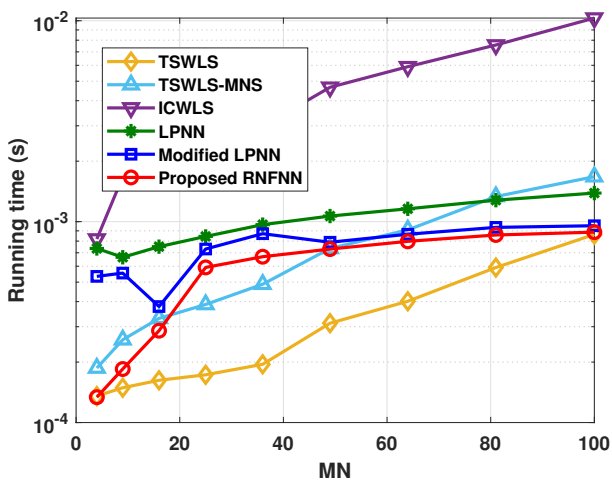}
	\caption{Running time versus $ M\!N $, where the proposed method is compared with the state-of-the-art algorithms. }
	\label{fig:runningtime}
\end{figure}

\textit{Scenario 4:} Based on the previous research studies (see e.g., \cite{2016-Khomchuk-Performanceanalysistarget} and \cite{2010-He-NoncoherentMIMORadar}), the localization performance depends on diversity gain $ M\!N $. In this scenario, we aim to investigate the positioning accuracy for radars with the different numbers of antennas and analyze the computational complexity of the proposed method via CPU running time averaged over 5000 Monte-Carlo trials. We set $ M = N $, and define $ R $ as $ 2000 m $. We place the target randomly within the 2-D area around the origin of a circle whose radius is $ R/2 $. The transmitters and the receivers are placed randomly on a circle whose radius is $ R $. In order to alleviate the effect due to the loss of geometric gain, we fix the positions of the 1st and 2nd transmitters at $ 0 $ and $ \pi $ , respectively, and the positions of the 1st and 2nd receivers at $ \pi/2 $ and $ -\pi/2 $, respectively (because when $ M\!N $ is too small, due to the loss of the geometric gain, the estimation performance becomes strongly dependent on the particular geometric scenario and can degrade significantly \cite{2016-Khomchuk-Performanceanalysistarget}). The SNR setup for this scenario is SNR = -20 dB. The CRLB is averaged over different geometries used for each $ M\!N $ value.

As illustrated in Fig. \ref{fig:difMN_rmse}, the CRLB and RMSE can be lowered by increasing the factor $ M\!N $, which is consistent with the results of previous studies \cite{2016-Khomchuk-Performanceanalysistarget} and \cite{2010-He-NoncoherentMIMORadar}. The proposed RNFNN and modified LPNN approaches are able to attain the CRLB for different localization geometries. The LPNN method can also reach the CRLB for this 2-D scenario. However, the TSWLS, TSWLS-MNS, and ICWLS cannot obtain desirable performance in this low SNR scenario.

Figure \ref{fig:runningtime} shows the runtime comparison results of all algorithms. This simulation is conducted using a computer with Intel(R) Core(TM) i7-9700K CPU @ 3.60GHz and MATLAB R2019a. Considering the parallel computing characteristics of the neural network methods, we utilize the computer's multi-core processor for parallel computing; however, there is a maximum of 8 parallel pools available due to the hardware (only eight cores) limitation. Despite this limitation, the proposed method still has an advantage in running time compared to other state-of-the-art methods (except TSWLS). Moreover, as indicated in the figure, the running time of the neural network methods grows slowly with the increase of $ M\!N $. In contrast, for those methods based on multi-stage WLS strategies, e.g., TSWLS, TSWLS-MNS, and ICWLS, their running time increases rapidly with increasing $ M\!N $, which is consistent with the foregoing theoretical analysis in Section \ref{subsec_complexity}.


\begin{figure}[!t]
	\centering
	\includegraphics[width=3.5in]{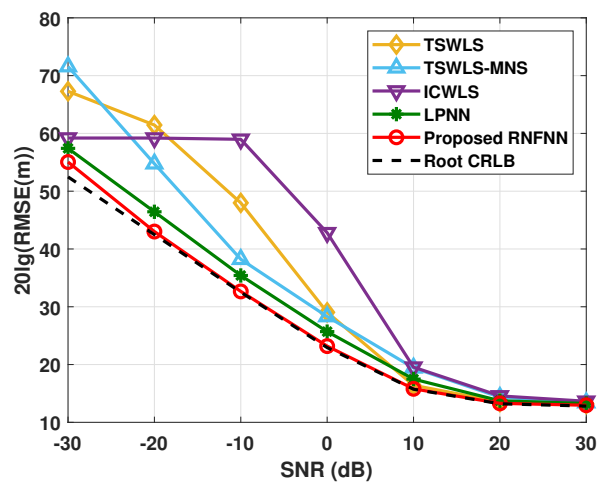}
	\caption{RMSE($ \hat{\boldsymbol{u}} $) versus SNR, where the proposed method is compared with the state-of-the-art algorithms and the root CRLB, in the case with antenna position errors.}
	\label{fig:3D_perror_snr}
\end{figure}

\begin{figure}[!t]
	\centering
	\includegraphics[width=3.5in]{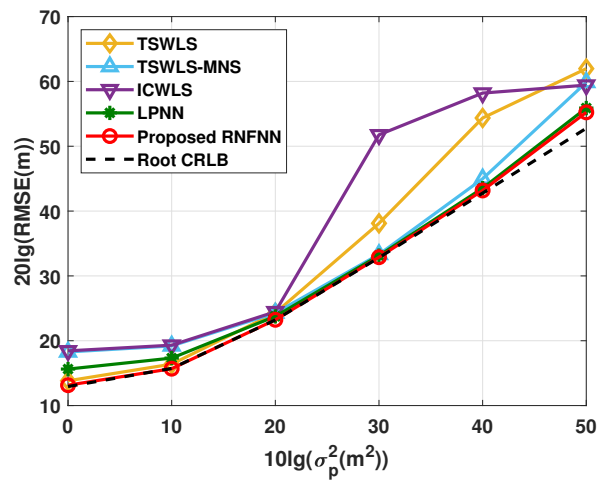}
	\caption{RMSE($ \hat{\boldsymbol{u}} $) versus the antenna position noise variance, where the proposed method is compared with the state-of-the-art algorithms, in the case with antenna position errors.}
	\label{fig:3D_perror_var}
\end{figure}

\subsection{Target Localization With Antenna Position Errors} \label{subsec:B}
In the simulations conducted here, the true positions of transmitter and receiver are not available, but we have their observed versions as in (\ref{eq:APmesurement_model}). For simplicity of analysis, the variance of the antenna position errors is assumed to be the same $ \sigma_{p}^2 = \sigma_{\boldsymbol{t},m}^2 = \sigma_{\boldsymbol{s},n}^2 $ for all $ m $ and $ n $, but it should be noted that the proposed method does not have such restriction. The noise-free version of antenna positions are tabulated in Table \ref{tab:3D_position}. The target is located at $ \boldsymbol{u}=[-500, 600, 550]^\text{T} m $. 
Note that TSWLS, TSWLS-MNS, and ICWLS do not consider the antenna position errors, but the extended version of the LPNN considers the antenna position errors.

\textit{Scenario 5:} In this scenario, we aim to evaluate the performance of the proposed method versus SNR. The antenna uncertainty setup is
considered as $ \sigma_{p}^2 = 10 m^2 $. In Fig. \ref{fig:3D_perror_snr}, we plot the RMSE of target location estimate as a function of SNR and compare it with the state-of-the-art methods and the CRLB. As shown in this figure, the proposed method can attain the CRLB performance up to relatively low SNR levels, which demonstrates the efficiency of the extended RNFNN method in present of antenna position errors.

\textit{Scenario 6:} In this scenario, we assess the performance of the proposed method versus the antenna uncertainty. The SNR setup for this scenario is considered as SNR = 10 dB. In Fig. \ref{fig:3D_perror_var}, we plot the positioning RMSE's of the localization methods as a function of noise variance and compare them with the CRLB. The result of this figure also corroborates the optimum performance of the proposed method by achieving the CRLB for different uncertainty levels, even though it is slightly worse at high antenna position noise level.

\section{Conclusion} \label{sec_conclusion}
In this work, we have investigated the issue of target localization in distributed MIMO radar with analog neural networks. The ML target localization  has a complicated objective function, and it is tricky to optimize directly. We transform the optimization problem into a more tractable one by introducing some auxiliary variables, two different energy functions are constructed by taking different strategies, and their corresponding neural networks, i.e., the RNFNN and modified LPNN, are defined and compared. The RNFNN has much simpler structure and faster convergence speed, compared with the LPNN counterpart. We also extend the RNFNN method to the case that the positions of transmitter and receiver are not exactly known. Simulation results demonstrate the efficiency of the proposed method by achieving the CRLB accuracy over high, moderate and relatively low SNR levels. Simulation results in various scenarios demonstrate that the proposed method outperforms the state-of-the-art algorithms in terms of performance and computational complexity.


\section*{Acknowledgment}
The authors would like to thank the anonymous reviewers for their constructive comments which helped to improve the quality of this manuscript.

\ifCLASSOPTIONcaptionsoff
  \newpage
\fi




\bibliographystyle{IEEEtran}
\bibliography{References}

\begin{thebibliography}{10}
\providecommand{\url}[1]{#1}
\csname url@samestyle\endcsname
\providecommand{\newblock}{\relax}
\providecommand{\bibinfo}[2]{#2}
\providecommand{\BIBentrySTDinterwordspacing}{\spaceskip=0pt\relax}
\providecommand{\BIBentryALTinterwordstretchfactor}{4}
\providecommand{\BIBentryALTinterwordspacing}{\spaceskip=\fontdimen2\font plus
\BIBentryALTinterwordstretchfactor\fontdimen3\font minus
  \fontdimen4\font\relax}
\providecommand{\BIBforeignlanguage}[2]{{%
\expandafter\ifx\csname l@#1\endcsname\relax
\typeout{** WARNING: IEEEtran.bst: No hyphenation pattern has been}%
\typeout{** loaded for the language `#1'. Using the pattern for}%
\typeout{** the default language instead.}%
\else
\language=\csname l@#1\endcsname
\fi
#2}}
\providecommand{\BIBdecl}{\relax}
\BIBdecl

\bibitem{2008-Xu-Targetdetectionparameter}
L.~Xu, J.~Li, and P.~Stoica, ``Target detection and parameter estimation for
  {MIMO} radar systems,'' \emph{IEEE Transactions on Aerospace and Electronic
  Systems}, vol.~44, no.~3, pp. 927--939, Jul. 2008.

\bibitem{2016-Khomchuk-Performanceanalysistarget}
P.~Khomchuk, R.~S. Blum, and I.~Bilik, ``Performance analysis of target
  parameters estimation using multiple widely separated antenna arrays,''
  \emph{IEEE Transactions on Aerospace and Electronic Systems}, vol.~52, no.~5,
  pp. 2413--2435, Oct. 2016.

\bibitem{2010-He-NoncoherentMIMORadar}
Q.~He, R.~S. Blum, and A.~M. Haimovich, ``Noncoherent {MIMO} radar for location
  and velocity estimation: {M}ore antennas means better performance,''
  \emph{IEEE Transactions on Signal Processing}, vol.~58, no.~7, pp.
  3661--3680, Jul. 2010.

\bibitem{2021-Sadeghi-TargetLocalizationGeometry}
M.~Sadeghi, F.~Behnia, R.~Amiri, and A.~Farina, ``Target localization geometry
  gain in distributed {MIMO} radar,'' \emph{IEEE Transactions on Signal
  Processing}, vol.~69, pp. 1642--1652, 2021.

\bibitem{2007-Li-MIMORadarColocated}
J.~Li and P.~Stoica, ``{MIMO} radar with colocated antennas,'' \emph{IEEE
  Signal Processing Magazine}, vol.~24, no.~5, pp. 106--114, Sep. 2007.

\bibitem{2008-Haimovich-MIMORadarWidely}
A.~M. Haimovich, R.~S. Blum, and L.~J. Cimini, ``{MIMO} radar with widely
  separated antennas,'' \emph{IEEE Signal Processing Magazine}, vol.~25, no.~1,
  pp. 116--129, 2008.

\bibitem{2007-Li-ParameterIdentifiabilityMIMO}
J.~Li, P.~Stoica, L.~Xu, and W.~Roberts, ``On parameter identifiability of
  {MIMO} radar,'' \emph{IEEE Signal Processing Letters}, vol.~14, no.~12, pp.
  968--971, Dec. 2007.

\bibitem{2021-Shi-ParameterIdentifiabilityDiversity}
J.~Shi, Z.~Yang, and Y.~Liu, ``On parameter identifiability of
  diversity-smoothing-based {MIMO} radar,'' \emph{IEEE Transactions on
  Aerospace and Electronic Systems}, pp. 1--1, 2021, early Access Article.

\bibitem{2019-Cheng-JointDesignTransmit}
Z.~Cheng, B.~Liao, Z.~He, J.~Li, and J.~Xie, ``Joint design of the transmit and
  receive beamforming in {MIMO} radar systems,'' \emph{IEEE Transactions on
  Vehicular Technology}, vol.~68, no.~8, pp. 7919--7930, Aug. 2019.

\bibitem{2006-Fishler-Spatialdiversityradars}
E.~Fishler, A.~Haimovich, R.~Blum, L.~Cimini, D.~Chizhik, and R.~Valenzuela,
  ``Spatial diversity in radars—models and detection performance,''
  \emph{IEEE Transactions on Signal Processing}, vol.~54, no.~3, pp. 823--838,
  Mar. 2006.

\bibitem{2010-Godrich-TargetLocalizationAccuracy}
H.~Godrich, A.~M. Haimovich, and R.~S. Blum, ``Target localization accuracy
  gain in {MIMO} radar-based systems,'' \emph{IEEE Transactions on Information
  Theory}, vol.~56, no.~6, pp. 2783--2803, Jun. 2010.

\bibitem{2012-Niu-TargetLocalizationTracking}
R.~Niu, R.~S. Blum, P.~K. Varshney, and A.~L. Drozd, ``Target localization and
  tracking in noncoherent multiple-input multiple-output radar systems,''
  \emph{IEEE Transactions on Aerospace and Electronic Systems}, vol.~48, no.~7,
  pp. 1466--1489, Apr. 2012.

\bibitem{2014-Rui-EllipticLocalizationPerformance}
L.~Rui and K.~C. Ho, ``Elliptic localization: Performance study and optimum
  receiver placement,'' \emph{IEEE Transactions on Signal Processing}, vol.~62,
  no.~18, pp. 4673--4688, Sep. 2014.

\bibitem{2009-Godrich-Targetlocalisationtechniques}
H.~Godrich, A.~M. Haimovich, and R.~S. Blum, ``Target localisation techniques
  and tools for multiple-input multiple-output radar,'' \emph{IET Radar Sonar
  \& Navigation}, vol.~3, no.~4, pp. 314--327, Aug. 2009.

\bibitem{2013-Dianat-TargetLocalizationusing}
M.~Dianat, M.~R. Taban, J.~Dianat, and V.~Sedighi, ``Target localization using
  least squares estimation for {MIMO} radars with widely separated antennas,''
  \emph{IEEE Transactions on Aerospace and Electronic Systems}, vol.~49, no.~4,
  pp. 2730--2741, Oct. 2013.

\bibitem{2015-Einemo-Weightedleastsquares}
M.~Einemo and H.~C. So, ``Weighted least squares algorithm for target
  localization in distributed {MIMO} radar,'' \emph{Signal Processing}, vol.
  115, pp. 144--150, Oct. 2015.

\bibitem{2015-Noroozi-TargetLocalizationBistatic}
A.~Noroozi and M.~A. Sebt, ``Target localization from bistatic range
  measurements in multi-transmitter multi-receiver passive radar,'' \emph{IEEE
  Signal Processing Letters}, vol.~22, no.~12, pp. 2445--2449, Dec. 2015.

\bibitem{2017-Amiri-AsymptoticallyEfficientTarget}
R.~Amiri, F.~Behnia, and H.~Zamani, ``Asymptotically efficient target
  localization from bistatic range measurements in distributed {MIMO} radars,''
  \emph{IEEE Signal Processing Letters}, vol.~24, no.~3, pp. 299--303, Mar.
  2017.

\bibitem{2020-Noroozi-ClosedFormSolution}
A.~Noroozi, M.~A. Sebt, S.~M. Hosseini, R.~Amiri, and M.~M. Nayebi,
  ``Closed-form solution for elliptic localization in distributed {MIMO} radar
  systems with minimum number of sensors,'' \emph{IEEE Transactions on
  Aerospace and Electronic Systems}, vol.~56, no.~4, pp. 3123--3133, Aug. 2020.

\bibitem{2017-Noroozi-IterativeTargetLocalization}
A.~Noroozi, A.~H. Oveis, and M.~A. Sebt, ``Iterative target localization in
  distributed {MIMO} radar from bistatic range measurements,'' \emph{IEEE
  Signal Processing Letters}, vol.~24, no.~11, pp. 1709--1713, Nov. 2017.

\bibitem{2016-Liang-LagrangeProgrammingNeural}
J.~Liang, C.~S. Leung, and H.~C. So, ``Lagrange programming neural network
  approach for target localization in distributed {MIMO} radar,'' \emph{IEEE
  Transactions on Signal Processing}, vol.~64, no.~6, pp. 1574--1585, Mar.
  2016.

\bibitem{2020SunMovingTargetLocalization}
T.~Sun, C.-X. Dong, Y.~Mao, and M.-M. Liu, ``Moving target localization in
  multiple-input multiple-output radar systems without the prior knowledge of
  measurement noise powers,'' \emph{IEEE Communications Letters}, vol.~24,
  no.~9, pp. 1957--1960, 2020.

\bibitem{2020SongApproximatelyEfficientEstimator}
H.~Song, G.~Wen, and L.~Zhu, ``An approximately efficient estimator for moving
  target localization in distributed mimo radar systems in presence of sensor
  location errors,'' \emph{IEEE Sensors Journal}, vol.~20, no.~2, pp. 931--938,
  2020.

\bibitem{2021SongTargetLocalizationClock}
H.~Song, G.~Wen, Y.~Liang, L.~Zhu, and D.~Luo, ``Target localization and clock
  refinement in distributed mimo radar systems with time synchronization
  errors,'' \emph{IEEE Transactions on Signal Processing}, vol.~69, pp.
  3088--3103, 2021.

\bibitem{2018-Zhang-AdaptiveProjectionNeural}
Y.~Zhang, S.~Chen, S.~Li, and Z.~Zhang, ``Adaptive projection neural network
  for kinematic control of redundant manipulators with unknown physical
  parameters,'' \emph{IEEE Transactions on Industrial Electronics}, vol.~65,
  no.~6, pp. 4909--4920, Jun. 2018.

\bibitem{2020-Li-CooperativeKinematicControl}
X.~Li, Z.~Xu, S.~Li, H.~Wu, and X.~Zhou, ``Cooperative kinematic control for
  multiple redundant manipulators under partially known information using
  recurrent neural network,'' \emph{IEEE Access}, vol.~8, pp. 40\,029--40\,038,
  2020.

\bibitem{2014-Yan-RobustModelPredictive}
Z.~Yan and J.~Wang, ``Robust model predictive control of nonlinear systems with
  unmodeled dynamics and bounded uncertainties based on neural networks,''
  \emph{IEEE Transactions on Neural Networks and Learning Systems}, vol.~25,
  no.~3, pp. 457--469, Mar. 2014.

\bibitem{2016-Yan-TubeBasedRobust}
Z.~Yan, X.~Le, and J.~Wang, ``Tube-based robust model predictive control of
  nonlinear systems via collective neurodynamic optimization,'' \emph{IEEE
  Transactions on Industrial Electronics}, vol.~63, no.~7, pp. 4377--4386, Jul.
  2016.

\bibitem{2016-Liu-$L_1$MinimizationAlgorithms}
Q.~Liu and J.~Wang, ``$l_{1}$-minimization algorithms for sparse signal
  reconstruction based on a projection neural network,'' \emph{IEEE
  Transactions on Neural Networks and Learning Systems}, vol.~27, no.~3, pp.
  698--707, Mar. 2016.

\bibitem{2019-Xu-DiscreteTimeProjection}
B.~Xu, Q.~Liu, and T.~Huang, ``A discrete-time projection neural network for
  sparse signal reconstruction with application to face recognition,''
  \emph{IEEE Transactions on Neural Networks and Learning Systems}, vol.~30,
  no.~1, pp. 151--162, Jan. 2019.

\bibitem{2017-Kobayashi-SymmetricComplexValued}
M.~Kobayashi, ``Symmetric complex-valued hopfield neural networks,'' \emph{IEEE
  Transactions on Neural Networks and Learning Systems}, vol.~28, no.~4, pp.
  1011--1015, Apr. 2017.

\bibitem{2016-Minemoto-RetrievalperformanceHopfield}
T.~Minemoto, T.~Isokawa, N.~Matsui, M.~Kobayashi, and H.~Nishimura, ``Retrieval
  performance of hopfield associative memory with complex-valued and
  real-valued neurons,'' in \emph{2016 International Joint Conference on Neural
  Networks (IJCNN)}, Jul. 2016, pp. 4133--4138.

\bibitem{2013-Li-dynamicneuralnetwork}
S.~Li and F.~Qin, ``A dynamic neural network approach for solving nonlinear
  inequalities defined on a graph and its application to distributed,
  routing-free, range-free localization of {WSN}s,'' \emph{Neurocomputing},
  vol. 117, pp. 72--80, Oct. 2013.

\bibitem{2018-Han-AugmentedLagrangeProgramminga}
Z.~Han, C.~S. Leung, H.~C. So, and A.~G. Constantinides, ``Augmented lagrange
  programming neural network for localization using time-difference-of-arrival
  measurements,'' \emph{IEEE Transactions on Neural Networks and Learning
  Systems}, vol.~29, no.~8, pp. 3879--3884, Aug. 2018.

\bibitem{2020-Shi-RobustMIMORadar}
Z.~Shi, H.~Wang, C.~S. Leung, H.~C. So, and M.~Eurasip, ``Robust {MIMO} radar
  target localization based on {Lagrange} programming neural network,''
  \emph{Signal Processing}, vol. 174, p. 107574, Sep. 2020.

\bibitem{1984-Barton-HandbookRadarMeasurement}
D.~K. Barton and H.~Ward, \emph{Handbook of {R}adar {M}easurement}.\hskip 1em
  plus 0.5em minus 0.4em\relax Norwood, MA, USA: Artech House, 1984.

\bibitem{2005-Willis-BistaticRadar}
J.~N. Willis, \emph{Bistatic {R}adar}.\hskip 1em plus 0.5em minus 0.4em\relax
  Raleigh, NC, USA: SciTech, 2005.

\bibitem{2013-Zhang-Matrixanalysisapplications}
X.~Zhang, \emph{Matrix {A}nalysis and {A}pplications}.\hskip 1em plus 0.5em
  minus 0.4em\relax Higher Education Press, 2013.

\bibitem{2019-Jia-EffectSensorMotion}
T.~Jia, K.~C. Ho, H.~Wang, and X.~Shen, ``Effect of sensor motion on time delay
  and doppler shift localization: {A}nalysis and solution,'' \emph{IEEE
  Transactions on Signal Processing}, vol.~67, no.~22, pp. 5881--5895, Nov.
  2019.

\bibitem{2020-Noroozi-EfficientClosedForm}
A.~Noroozi, R.~Amiri, M.~M. Nayebi, and A.~Farina, ``Efficient closed-form
  solution for moving target localization in {MIMO} radars with minimum number
  of antennas,'' \emph{IEEE Transactions on Signal Processing}, vol.~68, pp.
  2545--2557, 2020.

\bibitem{2015-Yi-MIMOPassiveRadar}
J.~Yi, X.~Wan, H.~Leung, and F.~Cheng, ``{MIMO} passive radar tracking under a
  single frequency network,'' \emph{IEEE Journal of Selected Topics in Signal
  Processing}, vol.~9, no.~8, pp. 1661--1671, Dec. 2015.

\bibitem{2021-Kazemi-DataAssociationMulti}
S.~A.~R. Kazemi, R.~Amiri, and F.~Behnia, ``Data association for multi-target
  elliptic localization in distributed {MIMO} radars,'' \emph{IEEE
  Communications Letters}, vol.~25, no.~9, pp. 2904--2907, Sep. 2021.

\bibitem{1988-Hopfield-Artificialneuralnetworks}
J.~Hopfield, ``Artificial neural networks,'' \emph{IEEE Circuits and Devices
  Magazine}, vol.~4, no.~5, pp. 3--10, Sep. 1988.

\bibitem{1992-Zhang-Lagrangeprogrammingneural}
S.~Zhang and A.~Constantinides, ``Lagrange programming neural networks,''
  \emph{IEEE Transactions on Circuits and Systems II: Analog and Digital Signal
  Processing}, vol.~39, no.~7, pp. 441--452, Jul. 1992.

\bibitem{2018-Han-AugmentedLagrangeProgramming}
Z.~Han, C.~S. Leung, H.~C. So, and A.~G. Constantinides, ``Augmented lagrange
  programming neural network for localization using time-difference-of-arrival
  measurements,'' \emph{IEEE Transactions on Neural Networks and Learning
  Systems}, vol.~29, no.~8, pp. 3879--3884, Aug. 2018.

\bibitem{1986-Tank-SimpleNeuralOptimization}
D.~Tank and J.~Hopfield, ``Simple 'neural' optimization networks: An {A/D }
  converter, signal decision circuit, and a linear programming circuit,''
  \emph{IEEE Transactions on Circuits and Systems}, vol.~33, no.~5, pp.
  533--541, May 1986.

\bibitem{2003-Leung-highperformancefeedback}
Y.~Leung, K.-Z. Chen, and X.-B. Gao, ``A high-performance feedback neural
  network for solving convex nonlinear programming problems,'' \emph{IEEE
  Transactions on Neural Networks}, vol.~14, no.~6, pp. 1469--1477, Nov. 2003.

\bibitem{1984HopfieldNeuronsgradedresponse}
J.~J. Hopfield, ``Neurons with graded response have collective computational
  properties like those of two-state neurons,'' \emph{Proceedings of the
  national academy of sciences}, vol.~81, no.~10, pp. 3088--3092, 1984.

\bibitem{1985-Hopfield-NeuralComputationDecisions}
J.~J. Hopfield, ``Neural computation of decisions in optimization problem,''
  \emph{Biological Cybernetics}, vol.~52, pp. 1119--1164, 1985.

\bibitem{2009-Lui-SemiDefiniteProgramming}
K.~W.~K. Lui, W.-K. Ma, H.~C. So, and F.~K.~W. Chan, ``Semi-definite
  programming algorithms for sensor network node localization with
  uncertainties in anchor positions and/or propagation speed,'' \emph{IEEE
  Transactions on Signal Processing}, vol.~57, no.~2, pp. 752--763, Feb. 2009.

\bibitem{2005-Patwari-Locatingnodescooperative}
N.~Patwari, J.~Ash, S.~Kyperountas, A.~Hero, R.~Moses, and N.~Correal,
  ``Locating the nodes: cooperative localization in wireless sensor networks,''
  \emph{IEEE Signal Processing Magazine}, vol.~22, no.~4, pp. 54--69, Jul.
  2005.

\bibitem{2005-Gustafsson-Mobilepositioningusing}
F.~Gustafsson and F.~Gunnarsson, ``Mobile positioning using wireless networks:
  possibilities and fundamental limitations based on available wireless network
  measurements,'' \emph{IEEE Signal Processing Magazine}, vol.~22, no.~4, pp.
  41--53, Jul. 2005.

\bibitem{2007HoSourceLocalizationUsing}
K.~C. Ho, X.~Lu, and L.~Kovavisaruch, ``Source localization using {TDOA} and
  {FDOA} measurements in the presence of receiver location errors: Analysis and
  solution,'' \emph{IEEE Transactions on Signal Processing}, vol.~55, no.~2,
  pp. 684--696, Feb. 2007.

\bibitem{2019AmiriEfficientEllipticLocalization}
R.~Amiri, S.~A.~R. Kazemi, F.~Behnia, and A.~Noroozi, ``Efficient elliptic
  localization in the presence of antenna position uncertainties and clock
  parameter imperfections,'' \emph{IEEE Transactions on Vehicular Technology},
  vol.~68, no.~10, pp. 9797--9805, Oct. 2019.

\bibitem{1993-Kay-FundamentalsStatisticalSignal}
S.~M. Kay, \emph{Fundamentals of {S}tatistical {S}ignal {P}rocessing:
  {E}stimation {T}heory}.\hskip 1em plus 0.5em minus 0.4em\relax Upper Saddle
  River, NJ, USA: Prentice-Hall, 1993.

\bibitem{2016-He-GeneralizedCramerRao}
Q.~He, J.~Hu, R.~S. Blum, and Y.~Wu, ``Generalized {Cramér–Rao} bound for
  joint estimation of target position and velocity for active and passive radar
  networks,'' \emph{IEEE Transactions on Signal Processing}, vol.~64, no.~8,
  pp. 2078--2089, Apr. 2016.

\bibitem{2013SoSimpleFormulaeBias}
H.~C. So, Y.~T. Chan, K.~Ho, and Y.~Chen, ``Simple formulae for bias and mean
  square error computation [{DSP} tips and tricks],'' \emph{IEEE Signal
  Processing Magazine}, vol.~30, no.~4, pp. 162--165, Jul. 2013.

\bibitem{2019WangConvexRelaxationMethods}
G.~Wang and K.~C. Ho, ``Convex relaxation methods for unified near-field and
  far-field {TDOA}-based localization,'' \emph{IEEE Transactions on Wireless
  Communications}, vol.~18, no.~4, pp. 2346--2360, Apr. 2019.

\end{thebibliography}

\bstctlcite{IEEEexample:BSTcontrol}

\end{document}